\documentclass[11pt]{article}
\usepackage{epsf}
\usepackage{color}

\usepackage{amsmath,amsfonts,amsbsy,amssymb}
\usepackage{indentfirst}
\usepackage{mathtools}

\newcommand{\fsl}[1]{\ensuremath{\mathrlap{\not{\phantom{#1}}}#1}}

\newcommand{\nn}{\nonumber}

\def\be{\begin{equation}}
\def\ee{\end{equation}}
\def\bse{\begin{subequations}}
\def\ese{\end{subequations}}
\def\bal{\begin{align}}
\def\ealn{\end{align}}
\def\tBF{\text{BF}}
\def\tr{\text{tr}}
\def\bs{\boldsymbol}

\begin{document}

\begin{titlepage}

\def\slash#1{{\rlap{$#1$} \thinspace/}}

\begin{flushright} 

\end{flushright} 

\vspace{0.1cm}

\begin{Large}
\begin{center}

{\bf    
Higher (Odd) Dimensional Quantum Hall Effect  
\\and \\
Extended Dimensional Hierarchy 
}
\end{center}
\end{Large}

\vspace{0.4cm}

\begin{center}
{\bf Kazuki Hasebe}   \\ 
\vspace{0.3cm} 
\it{Sendai National College of Technology, Ayashi, Sendai 989-3128, Japan } \\ 

\vspace{0.4cm} 
{\sf
khasebe@sendai-nct.ac.jp} 

\vspace{0.4cm} 

{\today} 

\end{center}

\vspace{0.2cm}

\begin{abstract}
\noindent

\baselineskip=18pt

We demonstrate dimensional ladder of higher dimensional quantum Hall effects by exploiting quantum Hall effects on arbitrary odd dimensional spheres. Non-relativistic and relativistic Landau models are analyzed on $S^{2k-1}$ in the $SO(2k-1)$ monopole background.  
The total  sub-band  degeneracy of the odd dimensional lowest Landau level  is shown to be equal to the winding number from the base-manifold $S^{2k-1}$ to  the one-dimension higher $SO(2k)$ gauge group. 
 Based on the chiral Hopf maps, we clarify the underlying  quantum Nambu geometry for odd dimensional quantum Hall effect and  the resulting  quantum geometry  is naturally embedded also  in one-dimension higher quantum geometry.   
An origin of such dimensional ladder connecting even and odd dimensional quantum Hall effects is illuminated  from a viewpoint of the spectral flow  of Atiyah-Patodi-Singer index theorem in differential topology.  
We also present a BF topological field theory as an effective field theory 
in which  membranes with different dimensions undergo non-trivial linking in odd dimensional space. Finally, an extended version of the dimensional hierarchy  for higher dimensional quantum Hall liquids is proposed, and its relationship to quantum anomaly and D-brane physics is discussed.

\baselineskip=18pt

\end{abstract}

\end{titlepage}

\newpage 

\tableofcontents

\newpage

\section{Introduction}\label{sec:intro}
 
 Topological insulators \cite{Hasan-Kane-10,Qi-Zhang-11} have attracted a great deal of attention in contemporary physics. Quantum Hall effect \cite{Laughlin1983, Haldane1983} is a  well investigated topological insulator  fitted in the 2D A-class of the periodic table   
(Table \ref{table:periodictable}) \cite{
Ryu-S-F-L-2010,SchnyderRFL2008,Kitaev2008}. Other A-class topological insulators which appear in arbitrary even dimensions can be regarded  as higher dimensional analogues of the 2D quantum Hall effect \cite{Hasebe-2014-1, HasebeKimura2003}. Zhang and Hu launched the study of the higher dimensional quantum Hall effect by their  four-dimensional system \cite{ZhangHu2001} and it has been further generalized by many researchers, for instance,  on complex projective manifolds \cite{KarabaliNair2002}.\footnote{There are so many works up to  the present. See Refs.\cite{Hasebe2010,  Karabali-Nair-2006} as reviews and references therein at that time.}  However most  works have  focused on even dimensional quantum Hall effects, and studies about odd dimensional quantum Hall effects are very few \cite{Nair&Daemi2004, Hasebe-2014-2}. As the even dimensional quantum Hall effect is a manifestation of the A-class topological insulator \cite{Hasebe-2014-1}, 
AIII class is naturally regarded as  odd dimensional counterpart of the quantum Hall effects \cite{Hasebe-2014-2}. 
A-class and AIII-class are special  topological insulators in the sense that  the topological invariant $\mathbb{Z}$ regularly appear in even and odd spacial dimensions in the periodic table of topological insulators (Table \ref{table:periodictable}).  
Indeed,  intimate relations are known to exist between A and AIII classes: AIII-class topological insulators can be induced by  dimensional reduction from one-dimension higher A-class topological insulators \cite{Ryu-S-F-L-2010}.

\begin{table}
\begin{center}
   \begin{tabular}{|c||c|c|c||c|c|c|c|c|c|c|c|c|c|}\hline
    A-Z    &  T &  C & TC & 1 & 2 & 3 & 4  & 5 & 6 & 7 & 8 \\ \hline \hline%
 A  &  $\circ$   & $\circ$   & $\circ$  & $\circ$ & $\mathbb{Z}$  & $\circ$ & $\mathbb{Z}$& $\circ$  & $\mathbb{Z}$&  $\circ$ & $\mathbb{Z}$                          \\ \hline %
 AIII &  $\circ$    & $\circ$  & 1 & $\mathbb{Z}$ & $\circ$ & $\mathbb{Z}$  & $\circ$  & $\mathbb{Z}$  & $\circ$ & $\mathbb{Z}$  &  $\circ$                  \\ \hline  \hline
 AI &  +1    &  $\circ$ & $\circ$ & $\circ$ & $\circ$ & $\circ$  & $\mathbb{Z}$  & $\circ$    & $\mathbb{Z}_2$ & $\mathbb{Z}_2$ & $\mathbb{Z}$                  \\ \hline 
 BDI &  +1    & +1  & 1 & $\mathbb{Z}$ & $\circ$ & $\circ$  & $\circ$  & $\mathbb{Z}$    & $\circ$ & $\mathbb{Z}_2$ &    $\mathbb{Z}_2$               \\ \hline 
 D &  $\circ$    & +1  & $\circ$ & $\mathbb{Z}_2$  & $\mathbb{Z}$ & $\circ$  & $\circ$  & $\circ$  & $\mathbb{Z}$    & $\circ$ &  $\mathbb{Z}_2$                  \\ \hline 
 DIII &  $-1$    & $+1$  & 1 & $\mathbb{Z}_2$ & $\mathbb{Z}_2$ & $\mathbb{Z}$  & $\circ$  & $\circ$    & $\circ$ & $\mathbb{Z}$ &    $\circ$                \\ \hline 
 AII &  $-1$    & $\circ$ & $\circ$ &  $\circ$ &  $\mathbb{Z}_2$ & $\mathbb{Z}_2$  & $\mathbb{Z}$  & $\circ$    & $\circ$ & $\circ$ &      $\mathbb{Z}$             \\ \hline 
 CII &  $-1$    & $-1$  & $1$ & $\mathbb{Z}$ & $\circ$ & $\mathbb{Z}_2$  & $\mathbb{Z}_2$  & $\mathbb{Z}$    & $\circ$ & $\circ$ &    $\circ$               \\ \hline 
C &  $\circ$    & $-1$  & $\circ$ & $\circ$ & $\mathbb{Z}$ & $\circ$ & $\mathbb{Z}_2$  & $\mathbb{Z}_2$    & $\mathbb{Z}$ & $\circ$ &      $\circ$             \\ \hline      
CI &  $+1$    & $-1$  &  $1$ & $\circ$ & $\circ$ & $\mathbb{Z}$  & $\circ$  & $\mathbb{Z}_2$    & $\mathbb{Z}_2$ & $\mathbb{Z}$  &  $\circ$                  \\ \hline         
    \end{tabular}       
\end{center}
\caption{The periodic table for  topological insulator \cite{Ryu-S-F-L-2010}. 
}
\label{table:periodictable}
\end{table}

Interestingly,  A and AIII-class topological insulators accommodate  quantum Nambu geometry \cite{Nambu1973,CurtrightZachos2003,DeBellisSS2010}  as  a consequence of the level projection \cite{Neupertetal2012,Esienneetal2012,Shiozaki-Fujimoto-2013}.  In string theory, the quantum Nambu geometry is considered to be relevant to the geometry of  M-theory. For instance, 
the fuzzy three-sphere  \cite{Guralnik&Ramgoolam2001,Ramgoolam2001,Ramgoolam2002,Papageorgakis&Ramgoolam2006} appears in a geometry of M2-M5 boundstate \cite{BasuHarvey2004}, and the 
 quantum Nambu bracket  plays a crucial role in the BLG formulation of M theory \cite{BaggerLambert2006,Gustavsson2007, BaggerLambert2007}.\footnote{See Ref.\cite{Ho-Matsuo-2016} as a recent good review.} 
It is rather fascinating that such exotic geometry emerges  in the context of topological insulators.\footnote{
Non-commutative geometries for AII class topological insulator and topological semi-metal are also reported in Refs.\cite{LiWu2013,Li-I-Y-W-2012,LiZhangWu2012} and Ref.\cite{Chu-1976}. }   
Also,  the non-commutative geometry has been practically applied to investigations of fractional quantum Hall effect (see Ref.\cite{Ezawaetal2003} for instance). 
However,  odd dimensional quantum Nambu geometry is known to   suffer from several pathological properties   such as  difficulties in quantization from the Nambu-Poisson bracket to commutator and violation of fundamental identity,  both of which even dimensional Nambu geometry does not sustain \cite{CurtrightZachos2003} .  A natural way to resolve such difficulties is to realize odd dimensional Nambu geometry as a subspace of the consistent even dimensional Nambu geometry. Based on the idea,  fuzzy three-sphere  is constructed as a narrow strip around the equator of one-dimension higher fuzzy four-sphere \cite{Jabbari2004,JabbariTorabian2005}.   See also  earlier works \cite{Mukunda-Sudarshan-1976, Bayen-Flato-1975}.

In this paper, we construct higher  odd dimensional quantum Hall effect. 
 In Ref.\cite{Hasebe-2014-1}, we investigated even  dimensional quantum Hall effect  and unveiled the underlying quantum Nambu geometry. 
Also we demonstrated exact correspondence between the zero-mode Landau level degeneracy  and  Chern number of  non-Abelian monopole bundle configuration, guaranteed by the Atiyah-Singer index theorem \cite{Dolan2003}.   
In  Ref.\cite{Hasebe-2014-2}, we investigated three dimensional quantum Hall effect  introduced by Nair and Daemi \cite{Nair&Daemi2004} and established its relation to  fuzzy three-sphere.   
The Nair and Daemi's  setup is a natural three dimensional counterpart of the Haldane's  quantum Hall effect on $S^2$ \cite{Haldane1983} and  Zhang and Hu's on $S^4$.    We combine our two preceding works \cite{Hasebe-2014-1,Hasebe-2014-2} to explore an entire picture of the higher dimensional quantum Hall effects.   
We start from  odd-dimensional Landau problem and clarify the associated non-commutative geometry.    Though our starting point is  completely odd-dimension, what is interesting is that  the odd dimensional Landau model ``knows'' one-dimension higher lowest Landau level structure.\footnote{Such structure has been observed in the context of non-commutative geometry \cite{JabbariTorabian2005}.}  
This is the heart of this paper. As we will discuss in detail,  there is  a dimensional ladder between  even and odd dimensions.   
Such ladder is an analogue of the celebrated dimensional ladder of anomalies in relativistic quantum field theory \cite{Stora-1983,Zumino-1983,Baulieu-1984, GaumeGinsparg1984, Sumitani-1984, AlvarezGaume-Pietra-Moore-1985} and, in differential topology, the one known as the spectral flow of the Atiyah-Patodi-Singer index theorem \cite{Atiyah-Patodi-Singer-1975-1,Atiyah-Patodi-Singer-1975-2,Atiyah-Patodi-Singer-1976}. 
The existence of the dimensional ladder  is also compatible with the ideas of the dimensional relation between A and AIII topological insulators and that of  non-commutative geometry stated above.   

This paper is organized as follows. In Sec.\ref{sect:mathbackground}, we introduce the chiral Hopf maps based on the original Hopf maps as  mathematical background of  odd dimensional quantum Hall effect.  
Sec.\ref{sec:generalnon-rel} discusses the non-relativistic Landau model on $S^{2k-1}$, and in Sec.\ref{sec:noncomgeo} the corresponding quantum Nambu geometry is elucidated.  The relativistic Landau problem on odd dimensional sphere is exploited in Sec.\ref{landaudiracodd}. We demonstrate equality between the total dimension of sub-bands of the lowest Landau level in odd dimension and that of  even dimension.  Sec.\ref{sec:atiyahpatodisinger} provides an  explanation of the dimensional ladder based on the spectral flow argument of  Atiyah-Patodi-Singer. 
As  effective field theory for the odd dimensional quantum Hall effect, BF topological field theory is presented in Sec.\ref{sec:KmatrixChern-SimonsTheory}.  In Sec.\ref{sec:dimensionalladder}, we discuss physical implications of the dimensional ladder in the higher dimensional quantum Hall effect. Sec.\ref{sec:summary} is devoted to summary and discussions. 

\section{Mathematical Background: The Chiral Hopf Maps }\label{sect:mathbackground}

The quantum Hall effect on even dimensional sphere is constructed based on the Hopf maps  \cite{HasebeKimura2003}: 
\be
S^{4k-1}~\overset{S^{2k-1}}{\longrightarrow}~S^{2k}. 
\ee
($k=1, 2$ and $4$ respectively correspond to the 1st, 2nd and 3rd Hopf maps.) By imposing one more condition on the total manifold, 
we introduce a reduced Hopf map which we shall call the chiral Hopf map: 
\be
S^{2k-1}_L\otimes S^{2k-1}_R ~\overset{S_D^{2k-1}}{\longrightarrow}~S^{2k-1}.   
\label{reductionfromgenerdHopf}
\ee
We take the $SO(2k)$ gamma matrices $(2^k\times 2^k$ matrices) as 
\be
\Gamma_{\mu}=\begin{pmatrix}
0 & i\bar{\gamma}_{\mu} \\
-i\gamma_{\mu} & 0 
\end{pmatrix}~~~~~~~(\mu=1,2, \cdots, 2k) 
\label{gammamu2k}
\ee
where 
\be
\gamma_{i}=\bar{\gamma}_i~~(i=1,2,\cdots, 2k-1),~~~~\gamma_{2k} =-\bar{\gamma}_{2k}=i1_{2^{k-1}}   
\ee
with the $SO(2k-1)$ gamma matrices $\gamma_i$
\footnote{For $k=1$, $\gamma_{i=1}=1$, for $k=2$, $\gamma_i=\sigma_i$ $(i=1,2,3)$ and for $k=4$, $\gamma_i=-i\lambda_i$ $(i=1,2,\cdots, 7)$ with real antisymmetric matrices $\lambda_i$ (see Ref.\cite{Hasebe2010} for details).}:  
\be
\{\gamma_i, \gamma_j\}=2\delta_{ij}. 
\ee
$\Gamma_{\mu}$ (\ref{gammamu2k}) can be expressed as 
\begin{equation}
\Gamma_i=\begin{pmatrix}
0 & i\gamma_i \\
-i\gamma_i & 0 
\end{pmatrix}~(i=1,2,\cdots,{2k-1}), ~~~\Gamma_{2k}=
\begin{pmatrix}
0 & \bs{1}_{2^{k-1}} \\
\bs{1}_{2^{k-1}} & 0 
\end{pmatrix}.  \label{so2kgammamatrices}
\end{equation}
The corresponding $SO(2k)$ generators are derived as 
\be
\Sigma_{\mu\nu}=-i\frac{1}{4}[\Gamma_{\mu}, \Gamma_{\nu}]
=\begin{pmatrix}
\sigma_{\mu\nu} & 0 \\
0 & \bar{\sigma}_{\mu\nu}  
\end{pmatrix}, 
\label{blockdiagso2kge} 
\ee
where each set of $\sigma_{\mu\nu}$ and $\bar{\sigma}_{\mu\nu}$  composes the $SO(2k)$ algebra of the Weyl representation:  
\be
\sigma_{\mu\nu}=-i\frac{1}{4}[\bar{\gamma}_{\mu}\gamma_{\nu}-\bar{\gamma}_{\nu}\gamma_{\mu}], ~~~~~~
\bar{\sigma}_{\mu\nu}=-i\frac{1}{4}[{\gamma}_{\mu}\bar{\gamma}_{\nu}-{\gamma}_{\nu}\bar{\gamma}_{\mu}], 
\ee
with  
\be
\sigma_{ij}=\bar{\sigma}_{ij}=-i\frac{1}{4}[\gamma^i, \gamma^j], ~~~~~~~\sigma_{i,2k}=-\bar{\sigma}_{i,2k}=\frac{1}{2}\gamma^i. 
\ee
$\sigma_{ij}$ $(i,j=1,2,\cdots 2k-1)$ are $SO(2k-1)$ generators.  
The chiral matrix $\Gamma_{2k+1}$ is given by 
\be
\Gamma_{2k+1}=(-i)^k\Gamma_1\Gamma_2\cdots\Gamma_{2k}=
\begin{pmatrix}
\bs{1}_{2^{k-1}} & 0 \\
0 & -\bs{1}_{2^{k-1}}
\end{pmatrix}. 
\ee
To represent the total manifold $S^{2k-1}_{L}\otimes S^{2k-1}_R$, we  introduce a spinor  
\be
 \Psi=\begin{pmatrix}
 \Psi_L \\
 \Psi_R
 \end{pmatrix}, 
 \label{lrpsichiral}
\ee 
subject to  
\begin{align}
&\Psi^{\dagger}\Psi=\bold{1}, \nn\\
&\Psi^{\dagger}\Gamma_{2k+1} \Psi=0,  \label{onemorecondhopfspinor}
\end{align}
or 
\be
{\Psi_L}^{\dagger}\Psi_L={\Psi_R}^{\dagger}\Psi_R=\frac{1}{2} .  
\ee
Using Eq.(\ref{lrpsichiral}), we realize the chiral Hopf map (\ref{reductionfromgenerdHopf})  as  
\begin{align}
\Psi~\longrightarrow~x_{\mu}=\Psi^{\dagger} \Gamma_{\mu}\Psi=-i{\Psi_R}^{\dagger}\gamma_{\mu}\Psi_L +i{\Psi_L}^{\dagger}\bar{\gamma}_{\mu}\Psi_R ~~~~~~~~(\mu=1,2,\cdots,2k), 
\label{chiralhopfmapexp}
\end{align}
where $x_{\mu}$ satisfy 
\be
\sum_{\mu=1}^{2k}x_{\mu}x_{\mu}=(\Psi^{\dagger}\Psi)^2-(\Psi^{\dagger}\Gamma_{2k+1}\Psi)^2=4({\Psi_L}^{\dagger}\Psi_L)^2({\Psi_R}^{\dagger}\Psi_R)^2=1 .  
\ee
$x_{\mu}$ thus denote coordinates of $S^{2k-1}$. 
Inverting the map (\ref{chiralhopfmapexp}), $\Psi$ can be expressed as\footnote{Eq.(\ref{chiralhopfmapexp}) suggests that $\Psi$  is a coherent state that satisfies 
\be
\sum_{\mu=1}^{2k}x_{\mu}\Gamma_{\mu}\Psi=\Psi. 
\label{coherentstaterel}
\ee
Indeed with the use of Eq.(\ref{so2kgammamatrices}) and (\ref{explicitso2khopf}), one may readily verify Eq.(\ref{coherentstaterel}). 
} 
\be
\Psi=\begin{pmatrix}
\Psi_L\\
\Psi_R
\end{pmatrix}=
\frac{1}{2\sqrt{1+x_{2k}}}
\begin{pmatrix}
(1+x_{2k}+ix_i\gamma_i) \\
(1+x_{2k}-ix_i\gamma_i)
\end{pmatrix}. \label{explicitso2khopf}
\ee
Notice that we can obtain the chiral Hopf spinor (\ref{explicitso2khopf}) also from the original Hopf spinor \cite{Hasebe2010} 
\be
\Psi_{\text{ori}}=\frac{1}{\sqrt{2(1+x_{2k+1}})}
\begin{pmatrix}
(1+x_{2k+1})\psi  \\
(x_{2k}-ix_{i}\gamma_i)\psi
\end{pmatrix}~~~~~(\psi^{\dagger}\psi=1),
\ee
by focusing on the equator of $S^{2k}$ 
\be
x_{2k+1}=0
\ee
and choosing a gauge 
\be
\psi= \frac{1}{\sqrt{2(1+x_{2k})}} (1+x_{2k}+ix_i\gamma_i). 
\ee
In this sense, the chiral Hopf spinor is equivalent to the original Hopf spinor on the equator of $S^{2k}$.   

The chiral Hopf map naturally induces a product of  two Hopf maps: 
\be
S^{2k-1}_L\otimes S^{2k-1}_R~\overset{S^{k-1}_L\otimes S^{k-1}_R}{\longrightarrow}~S^k_L\otimes S^k_R. 
\ee
When $k=2, 4$ and $8$, $L$ and $R$ respectively correspond to 1st, 2nd and 3rd Hopf maps. They are explicitly given by  
\begin{align}
&\Psi_L~\rightarrow~x^i_L=\Psi_L^{\dagger}\gamma_i \Psi_L~~~~~(i=1,2,\cdots, 2k-1),\nn\\
&\Psi_R~\rightarrow~x^i_R=\Psi_R^{\dagger}\gamma_i \Psi_R~~~~~(i=1,2,\cdots, 2k-1).  
\end{align}
From $\Psi_L$ and $\Psi_R$, we can derive two gauge fields for $L$ and $R$:  
\begin{align}
&A_L=-i{\Psi_L}^{\dagger}d\Psi_L=-\frac{1}{2(1+x_{2k})}\bar{\sigma}_{\mu\nu}x_{\nu}dx_{\mu}+\frac{1}{4(1+x_{2k})}\gamma_idx_i,
\nn\\
&A_R=-i{\Psi_L}^{\dagger}d\Psi_L=-\frac{1}{2(1+x_{2k})}{\sigma}_{\mu\nu}x_{\nu}dx_{\mu}-\frac{1}{4(1+x_{2k})}\gamma_idx_i,
\end{align}
and then the total gauge field is given by 
\be
A=-i\Psi^{\dagger}d\Psi=A_L+A_R=A_{\mu} dx_{\mu} 
\label{so2k-1gaugefield}
\ee
with 
\be
A_i=-\frac{1}{1+x_{2k}}\sigma_{ij}x_j, 
~~~A_{2k}=0.  
\label{so2k-1gaugemonopolefield}
\ee
Notice that both $A_{L}$ and $A_{R}$ depend on the $SO(2k)$ generators but their sum amounts to  $A_{\mu}$ (\ref{so2k-1gaugefield}) that depends on the $SO(2k-1)$ generators only, and $A$  represents the $SO(2k-1)$ monopole gauge field.\footnote{ 
The $SO(2k-1)$ gauge field (\ref{so2k-1gaugemonopolefield}) exhibits singularity at the south pole on  $S^{2k-1}$. The $SO(2k-1)$ gauge field with  singularity at the north pole 
\be
A'=-\frac{1}{1-x_{2k}}\sigma_{ij}x_jdx_i
\ee
is related to  $A$ (\ref{so2k-1gaugemonopolefield}) by the following gauge transformation 
\be
A'=g^{\dagger}Ag-ig^{\dagger}dg 
\ee
with 
\be
g=\frac{1}{\sqrt{1-{x_{2k}}^2}} ix_i\gamma_i=-g^{\dagger}=-g^{-1}. 
\ee
Notice 
\be
A=-i\frac{1}{2}(1-x_{2k})dg g^{\dagger}, ~~~~
A'=-i\frac{1}{2}(1+x_{2k})g^{\dagger}dg.
\ee
The field strengths $F=dA+iA^2$ and $F'=dA'+i{A'}^2$ are related by $F'=g^{\dagger}Fg$.  
 }  
 The $SO(2k-1)$ monopole field strength $F_{\mu\nu}$ 
\be
F_{\mu\nu}=\partial_{\mu}A_{\nu}-\partial_{\nu}A_{\mu}+i[A_{\mu},A_{\nu}]
\ee
is obtained as 
\be
F_{ij}=-x_i A_j+x_jA_i+\sigma_{ij}, ~~~~F_{i, 2k}=(1+x_{2k})A_i. 
\ee
While the above derivation of $SO(2k-1)$ monopole gauge field  is apparently parallel to that of $SO(2k)$ gauge field from the Hopf map \cite{HasebeKimura2003},  there is a problem, because there does not exist homotopy that guarantees the quantization of the $SO(2k-1)$ monopole charge:  
\be
\pi_{2k-2}(SO(2k-1))=0. 
\ee
This problem is resolved when the present system is embedded in one-dimension higher system in which the monopole gauge field is $SO(2k)$ and then the charge is quantized:  
\be
\pi_{2k-1}(SO(2k))=\mathbb{Z}. 
\ee
 We will discuss the embedding later, but for the moment, we just suppose quantization of the $SO(2k-1)$ monopole charge to be quantized. 

\section{Non-Relativistic Landau Model on $S^{2k-1}$ }\label{sec:generalnon-rel}

Our next task is to analyze the non-relativistic Landau model on $S^{2k-1}$ in the $SO(2k-1)$ monopole background\footnote{In $2k$D, the square of the covariant derivative is given by 
\be
\sum_{\mu=1}^{2k}{D_{\mu}}^2= \frac{\partial^2}{\partial r^2} +(2k-1)\frac{1}{r}\frac{\partial}{\partial r} -\frac{1}{r^2} \sum_{\mu <\nu}{\Lambda_{\mu\nu}}^2.
\ee
},  
\be
H=-\frac{1}{2M}\sum_{\mu=1}^{2k}{D_{\mu}}^2\biggr|_{r=1}=\frac{1}{2M}\sum_{\mu<\nu =1}^{2k}{\Lambda_{\mu\nu}}^2, 
\label{so2k-1hamlannonrel}
\ee
where $\Lambda_{\mu\nu}$ are the $SO(2k)$ covariant angular momentum  
\be
\Lambda_{\mu\nu}=-ix_\mu D_\nu+ix_\nu D_\mu
\ee
with 
\be
D_\mu=\partial_\mu+iA_\mu. 
\ee
The total angular momentum is constructed as 
\be
L_{\mu\nu}=\Lambda_{\mu\nu}+F_{\mu\nu}. 
\label{so2ktotalangularmso2k}
\ee
In detail, they are 
\be
L_{ij}=L_{ij}^0 +\sigma_{ij}, ~~~~L_{i, 2k}=L_{i, 2k}^0 -\frac{1}{1+x_{2k}} \sigma_{ij}x_j  
\ee
with 
\be
L_{ij}^0=-ix_i\frac{\partial}{\partial x_j}+ix_j\frac{\partial}{\partial x_i}. 
\ee
It is easy to check that $L_{\mu\nu}$ satisfy the $SO(2k)$ algebra, and the $SO(2k)$ Casimir ${L_{\mu\nu}}^2$ is given by 
\be
{L_{\mu\nu}}^2={\Lambda_{\mu\nu}}^2+{F_{\mu\nu}}^2, 
\ee
and the $SO(2k)$ Landau Hamiltonian (\ref{so2k-1hamlannonrel}) can be rewritten as 
\be
H=\frac{1}{2M}(\sum_{\mu<\nu =1}^{2k}{L_{\mu\nu}}^2-\sum_{\mu<\nu =1}^{2k}{F_{\mu\nu}}^2)=
\frac{1}{2M}(C_{SO(2k)}-{C_{SO(2k-1)}}). 
\label{so2khamilLL}
\ee
Here we generalize the previous $SO(4)$ analysis \cite{Hasebe-2014-2} to  $SO(2k)$ $(k\ge 2)$ and specify the irreducible representation  as  
\be
[l_1,l_2,\cdots,l_{k-1},l_k]=[n+\frac{I}{2},\frac{I}{2},\cdots,\frac{I}{2}, s] \equiv [n, \frac{I}{2}, s]_{SO(2k)}, 
\label{so2knislong}
\ee
where  $n(=0, 1, 2, \cdots)$ denotes the Landau level index, and $s$ which we call the chirality parameter takes the following values, 
\be
s~=~\frac{I}{2},~\frac{I}{2}-1,~\cdots, ~-\frac{I}{2}+1, ~-\frac{I}{2}. 
\ee
Notice that the minimum of the absolute value of the chirality parameter takes different values   depending on the parity of $I$: For even $I$  0, while for odd $I$ $1/2$.  
The $SO(2k)$ Casimir is evaluated as\footnote{One can find the Casimir eigenvalues and the dimensions of the corresponding representations in Ref.\cite{IachelloBook}.} 
\be
C[n, \frac{I}{2}, \pm s]_{SO(2k)}=\sum_{i=1}^k l_i(l_i+2k-2i)=n(n+2k-2)+\frac{1}{2}I(2n+k^2-k)+\frac{1}{4}I^2 (k-1)+s^2 . 
\ee
For the $SO(2k-1)$ monopole gauge group, the representation  is given by  the fully symmetric representation  
\be
[l_1,l_2,\cdots,l_{k-1}]=[\frac{I}{2},\frac{I}{2},\cdots,\frac{I}{2}] \equiv [0, \frac{I}{2}]_{SO(2k-1)} , 
\ee
and the corresponding $SO(2k-1)$ Casimir eigenvalue becomes  
\be 
C[0, \frac{I}{2}]_{SO(2k-1)}=\sum_{i=1}^{k-1} l_i(l_i+2k-1-2i)=\frac{1}{4}I^2 (k-1)+\frac{1}{2}I(k-1)^2. 
\ee
Consequently, the energy eigenvalues  (\ref{so2khamilLL}) for $k \ge 2$ are derived as\footnote{Recovering the radius of sphere $r$ in the $SO(2k)$ Landau Hamiltonian (\ref{so2k-1hamlannonrel}) with $M\rightarrow Mr^2$, we take  the planar limit $I, r  ~\rightarrow~\infty$ with fixed $I/r^2$ and $s$. The $SO(2k)$ Landau levels (\ref{energyeiodddimension}) are reduced to 
\be
E_n~\rightarrow ~\frac{I}{2M}(n+\frac{k-1}{2}).  
\ee
These are equal to the planar limit of the Landau levels of the $SO(2k-1)$ Landau model \cite{Hasebe-2014-1}.} 
\be
E_n(s)=\frac{1}{2M}\biggl(n(n+2k-2)+{I}(n+\frac{k-1}{2})+s^2\biggr) =E_n(-s) . 
\label{energyeiodddimension}
\ee
Notice that the energy (\ref{energyeiodddimension}) depends not only on the Landau level index $n$ but also on the chirality parameter $s$.     
In particular  the ``lowest Landau level'' ($n=0$)  is split to  the sub-bands  
\be
E_{n=0}(s)=\frac{1}{2M}\biggl(\frac{I}{2}(k-1)+s^2\biggr) . 
\label{LLLenergyeiodddimension}
\ee
For $k=1$, the energy eigenvalues are given by (see Appendix \ref{sec:reviewcircle} for details) 
\begin{align}
\hspace{-1.2cm}SO(2)~~:~~&E_n =\frac{1}{2M} n^2 ~~~~~~~~~~~~(\text{for}~\text{even}~ I), \nn\\
&E_n =\frac{1}{2M} (n+\frac{1}{2})^2 ~~~~(\text{for}~\text{odd}~ I). 
\end{align}
 The corresponding degeneracy of the sub-band of the Landau level (\ref{energyeiodddimension}) is derived as  
\begin{align}
d[n, \frac{I}{2}, s]_{SO(2k)}&=\frac{(2n+I+2k-2)^2-4s^2}{4(k-1)^2}\cdot \prod_{2\le i \le k-1} \frac{(n+I+2k-i-1)(n+i-1)}{(2k-i-1)(i-1)}\nn\\
&\cdot \prod_{2 \le i < j \le k-1 } \frac{I+2k-i-j}{2k-i-j} \cdot \prod_{2 \le i \le k-1} \frac{(I+2k-2i)^2-4s^2}{4(k-i)^2} , 
\label{degelandaunlambda}
\end{align}
and for the lowest Landau level (\ref{LLLenergyeiodddimension}), the sub-band  
has the following degeneracy 
\be
d[0, \frac{I}{2}, s]_{SO(2k)}= d[0, \frac{I}{2}, -s]_{SO(2k)}=\prod_{j=1}^{k-1} \frac{(I+2j)^2-4s^2}{(2j)^2}  \times \prod_{l=1}^{k-2}\prod_{i=1}^{k-l-1}\frac{I+2l+i}{2l+i}.
\label{n=0numberofstates}
\ee
For instance\footnote{In particular, 
\be
d[0, 0, 0]_{SO(2k)}=1, ~~~~
d[0, \frac{1}{2}, \pm \frac{1}{2}]_{SO(2k)}= 2^{k-1}. 
\ee
The second relation stands for the well known formula of the dimension of the $SO(2k)$ fundamental spinor.  
}, 
\begin{align}
&d[0, \frac{I}{2},  s]_{SO(2)}\equiv 1~\sim ~I^0, \nn\\
&d[0, \frac{I}{2},  s]_{SO(4)}= \frac{1}{4}((I+2)^2-4s^2)~\sim ~I^2, \nn\\
&d[0, \frac{I}{2}, s]_{SO(6)}=\frac{1}{192}((I+2)^2-4s^2)(I+3)((I+4)^2-4s^2)~\sim ~I^5, \nn\\
&d[0,  \frac{I}{2},  s]_{SO(8)}=\frac{1}{138240}((I+2)^2-4s^2)(I+3)((I+4)^2-4s^2)(I+4)(I+5)((I+6)^2-4s^2)~\sim ~I^9. 
\label{lowdimexdso2k}
\end{align}
Here $\sim$ means the limit $I \rightarrow\infty$ with finite $s$.   
The chiral Hopf spinor (\ref{explicitso2khopf}) is the lowest Landau level eigenstate of the Landau Hamiltonian (\ref{so2k-1hamlannonrel}) for $\frac{I}{2}=|s|=\frac{1}{2}$. Taking the total antisymmetric combination of  $N$ chiral Hopf spinors, we can readily construct a Laughlin like $N$-body wave-function or Laughlin-Halperin like wave-function as in Ref.\cite{Hasebe-2014-2}. 

From Eq.(\ref{n=0numberofstates}), one may find a hierarchical relation in the degeneracies between different dimensions ($2k-1$ and $2k-3$): 
\begin{align}
d[0, \frac{I}{2}, s]_{SO(2k)}
&=\frac{(k-1)!}{(2k-3)!}\frac{(I+2k-2)^2-4s^2}{(2k-2)^2}\frac{(I+2k-3)!}{(I+k-1)!}~\cdot~ d[0, \frac{I}{2}, s]_{SO(2k-2)}\nn\\
&~\sim~{I}^{k}~\cdot~  d[0, \frac{I}{2}, s]_{SO(2k-2)}. 
\end{align}
The total dimension of the lowest Landau level can be read off from Eq.(\ref{n=0numberofstates}):  
\be
\sum_{s= -\frac{I}{2}}^{\frac{I}{2}} d[0, \frac{I}{2}, s]_{SO(2k)}=\prod_{l=1}^k\prod_{i=1}^l\frac{I+l+i-1}{l+i-1}. 
\label{sumlambdadege}
\ee
For instance, 
\begin{align}
&\sum_{s= -\frac{I}{2}}^{\frac{I}{2}} d[0, \frac{I}{2}, s]_{SO(2)}=I+1, \nn\\
&\sum_{s= -\frac{I}{2}}^{\frac{I}{2}} d[0, \frac{I}{2}, s]_{SO(4)}=\frac{1}{6}(I+1)(I+2)(I+3), \nn\\
&\sum_{s= -\frac{I}{2}}^{\frac{I}{2}} d[0, \frac{I}{2}, s]_{SO(6)}=\frac{1}{360}(I+1)(I+2)(I+3)^2(I+4)(I+5), \nn\\
&\sum_{s= -\frac{I}{2}}^{\frac{I}{2}} d[0, \frac{I}{2}, s]_{SO(8)}=\frac{1}{302400}(I+1)(I+2)(I+3)^2(I+4)^2(I+5)^2(I+6)(I+7). 
\label{sum2kexapdege}
\end{align}
Interestingly, these quantities correspond to the winding numbers from $S^{2k-1}$ to $SO(2k)$ gauge group (rather than the original $SO(2k-1)$ monopole gauge group):  
\be
\pi_{2k-1}(SO(2k))~\simeq ~\mathbb{Z},  
\label{mapfrom2k-1to2k}
\ee
which are represented as 
\be
\nu_{k} = \frac{1}{(2k-1)!(2i)^{k-1}\mathcal{A}(S^{2k-1})} \int_{S^{2k-1}} \tr (-ig^{\dagger}dg )^{2k-1} =(-i)^{k-1} \frac{1}{(2\pi)^k} \frac{(k-1)!}{(2k-1)!} \int_{S^{2k-1}} \tr (-ig^{\dagger}dg )^{2k-1}.  
\label{windingnumberexpli}
\ee
Here  $\mathcal{A}(S^{2k-1})=\frac{2\pi^k}{(k-1)!}$ is the area of $S^{2k-1}$, and  $g$ denotes the $SO(2k)$ group element.  When $g$ is the $SO(2k)$ element of the fully symmetric representation $[0, \frac{I}{2}, \frac{I}{2}]_{SO(2k)}$, 
 the winding number is explicitly evaluated as  (see Appendix \ref{append:windingso2k}) 
\be
\nu_k(I) 
=\frac{1}{2^{k-1}} \frac{(I+2k-4)!!}{(2k-3)!! I!!} \sum_{s=-\frac{I}{2}}^{\frac{I}{2}}(2s)^2 d[0,\frac{I}{2}, s]_{SO(2k-2)}~~~~(k\ge 2), 
\label{fomulaforwindingexp}
\ee
 and then\footnote{For $\nu_1(I)$, we used $g=(\frac{1}{\sqrt{1-{x_3}^2}} (x_2-ix_1))^I=(-i)^Ie^{iI\phi}$.} 
\begin{align}
&\nu_1(I)=\frac{1}{2\pi}\int_{S^1}\tr(-ig^{\dagger}dg)=I, ~~~~~~~~~~\nn\\
&\nu_2(I)=-i\frac{1}{24\pi^2}\int_{S^3}\tr(-ig^{\dagger}dg)^3=\frac{1}{6}I(I+1)(I+2), \nn\\
&\nu_3(I)=-\frac{1}{480\pi^3}\int_{S^5}\tr(-ig^{\dagger}dg)^5=\frac{1}{360}I(I+1)(I+2)^2(I+3)(I+4), \nn\\
&\nu_4(I)=i\frac{1}{13440\pi^4}\int_{S^7}\tr(-ig^{\dagger}dg)^7=\frac{1}{302400}I(I+1)(I+2)^2(I+3)^2(I+4)^2(I+5)(I+6),   
\label{windingnumbersk}
\end{align}
where we used Eq.(\ref{lowdimexdso2k}) to evaluate $\sum_{s=-\frac{I}{2}}^{\frac{I}{2}}(2s)^2 d[0,\frac{I}{2}, s]_{SO(2k-2)}~$. 
Comparison between Eq.(\ref{sum2kexapdege}) and (\ref{windingnumbersk}) manifests  the exact equality between the total number of the $(2k-1)$D lowest Landau level states for $I$ and the $SO(2k)$ winding number for $I+1$: 
\be
\sum_{s= -\frac{I}{2}}^{\frac{I}{2}}  d[0, \frac{I}{2}, s]_{SO(2k)}  =\nu_k(I+1). 
 \label{indextheoremvariant}
\ee
The origins of the  $SO(2k)$ gauge group and the relation (\ref{indextheoremvariant}) will be unveiled in the following sections both  from   non-commutative geometry and  differential topology point of view.  

\section{Non-Commutative Geometry}\label{sec:noncomgeo}

 We now discuss non-commutative geometry associated with the odd dimensional Landau model. Quantization of the chiral Hopf maps naturally realizes the geometry of  odd dimensional fuzzy spheres.

\subsection{Quantum Nambu geometry}

In the lowest Landau level the complex conjugate is replaced with derivative, 
\be
\Psi^{*}~\rightarrow ~\frac{\partial}{\partial \Psi}, 
\ee
and the coordinates on $S^{2k-1}$ introduced by the chiral Hopf map (\ref{chiralhopfmapexp}) are transformed to the operators
\be
X_{\mu}= \alpha \Psi^t \Gamma_{\mu} \frac{\partial}{\partial \Psi} 
\label{defcoors2k-1}
\ee
or 
\be
X_{\mu}=i\alpha \Psi^t_L \bar{\gamma}_{\mu}\frac{\partial}{\partial \Psi_R} -i\alpha \Psi^t_R {\gamma}_{\mu}\frac{\partial}{\partial \Psi_L}, 
\ee
where $\Gamma_{\mu}$ are the $SO(2k)$ gamma matrices (\ref{gammamu2k}), and  $\alpha$ denotes the parameter of the non-commutative scale;  
\be
\alpha=\frac{2r}{I}. 
\ee
(We recovered the radius of sphere $r$.) 
The commutation relation of  $X_{\mu}$ yields\footnote{
Expanding the $SO(2k)$ generators by the $Spin(2k)$ generators, $t^A$ $(A=1, 2, \cdots, k(2k-1))$,  
\be
\sigma_{\mu\nu}^+=\sum_{A=1}^{k(2k-1)}  \eta_{\mu\nu}^A t_A,~~~~~{\sigma}_{\mu\nu}^-=\sum_{A=1}^{k(2k-1)}  \bar{\eta}_{\mu\nu}^A t_A, 
\ee
we can express Eq.(\ref{commxx}) as 
\be
[X_{\mu}, X_{\nu}]=4i\alpha  \sum_{A=1}^{k(2k-1)} (\eta_{\mu\nu}^A X_A^L +\bar{\eta}_{\mu\nu}^A X_A^R) 
\ee
with
\be
X_{A}^L\equiv \alpha \Psi_L^t~ t_A \frac{\partial}{\partial \Psi_L},~~~X_{A}^R\equiv \alpha \Psi_R^t~ t_A \frac{\partial}{\partial \Psi_R}. 
\ee
For $SO(4)$, the expansion coefficients $\eta_{\mu\nu}^A$ and  $\bar{\eta}_{\mu\nu}^A$ are self- and anti-self dual 'tHooft symbols. For  $SO(2k)$, $\sigma_{\mu\nu}^{+}$ and $\sigma_{\mu\nu}^-$ satisfy the generalized self- and anti-self dual relations respectively: 
\be
\sigma^{\pm}_{\mu_1\mu_2} =\pm \frac{2^{k-2}}{(2k-2)!}\epsilon_{\mu_1\mu_2\cdots\mu_{2k-1}\mu_{2k}}\sigma^{\pm}_{\mu_3\mu_4}\cdots \sigma^{\pm}_{\mu_{2k-1}\mu_{2k}}. 
\ee
} 
\be
[X_{\mu}, X_{\nu}]=4i\alpha (X^L_{\mu\nu} +{X}^R_{\mu\nu}),  
\label{commxx}
\ee
where 
\be
X_{\mu\nu}^L\equiv \alpha \Psi_L^t ~\sigma_{\mu\nu}^+ \frac{\partial}{\partial \Psi_L},~~~X_{\mu\nu}^R\equiv \alpha \Psi_R^t ~{\sigma}_{\mu\nu}^- \frac{\partial}{\partial \Psi_R}.
\ee
Notice that only among $X_{\mu}$, their commutators do not close but introduce the new operators,  
$X_{\mu\nu}^L$ and $X_{\mu\nu}^R$, that  respectively satisfy the independent $SO(2k)$ algebra: 
\begin{align}
&[X^L_{\mu\nu}, X^L_{\rho\sigma}]= i \alpha \delta_{\mu\rho}X_{\nu\sigma}^L-i\alpha \delta_{\mu\sigma}X^L_{\nu\rho}+i\alpha \delta_{\nu\sigma}X^L_{\mu\rho}-i\alpha \delta_{\nu\rho}X^L_{\mu\sigma}, \nn \\
&[X^R_{\mu\nu}, X^R_{\rho\sigma}]= i\alpha \delta_{\mu\rho}X_{\nu\sigma}^R-i\alpha \delta_{\mu\sigma}X^R_{\nu\rho}+i\alpha \delta_{\nu\sigma}X^R_{\mu\rho}-i\alpha \delta_{\nu\rho}X^R_{\mu\sigma}, \nn\\
&[X^L_{\mu\nu}, X^R_{\rho\sigma}]=0. 
\end{align}
Furthermore, with 
\be
X_{\mu\nu}\equiv X_{\mu\nu}^L+X_{\mu\nu}^R, 
\ee
we have 
\begin{align}
&[X_{\mu}, X_{\nu}]=4i\alpha X_{\mu\nu}, \nn\\
&[X_{\mu\nu}, X_{\rho}]=i\delta_{\mu\rho}X_{\nu}-i\delta_{\nu\rho}X_{\mu}, \nn\\
&[X_{\mu\nu}, X_{\rho\sigma}]= i\delta_{\mu\rho}X_{\nu\sigma}-i\delta_{\mu\sigma}X_{\nu\rho}+i\delta_{\nu\sigma}X_{\mu\rho}-i\delta_{\nu\rho}X_{\mu\sigma}. 
\label{commuxss}
\end{align}
With the identification $X_{\mu, 2k+1}=- X_{2k+1, \mu}\equiv \frac{1}{2}X_{\mu}$, Eq.(\ref{commuxss}) is given by the closed $SO(2k+1)$ algebra  of $X_{ab}$ $(a,b=1,2,\cdots, 2k+1)$ determining the non-commutative structure of $S_F^{2k-1}$. 

Instead of using the usual commutator formalism,  we can describe $S_F^{2k-1}$ in quantum Nambu formalism where   
 $X_{\mu}$ satisfy a concise algebra.   
The quantum Nambu bracket or $n$-bracket is defined by   
\be
[O_{a_1}, O_{a_2}, \cdots, O_{a_n}]\equiv O_{[a_1}O_{a_2\cdots }O_{a_n]}, 
\label{usualdefnambu}
\ee
where the right-hand side represents the totally antisymmetric combination of the $n$ quantities.   For instance, 
\begin{align}
&[O_{a_1}, O_{a_2}]=O_{a_1} O_{a_2}-O_{a_2} O_{a_1}, \nn\\
&[O_{a_1}, O_{a_2}, O_{a_3}] =O_{a_1} O_{a_2} O_{a_3} -O_{a_1} O_{a_3} O_{a_2} - O_{a_2} O_{a_1} O_{a_3} +  O_{a_2} O_{a_3} O_{a_1} + O_{a_3} O_{a_1} O_{a_2} - O_{a_3} O_{a_2} O_{a_1}.  
\end{align}
Unlike the case of even D fuzzy spheres \cite{Hasebe-2014-1},  the quantized coordinates (\ref{defcoors2k-1}) do not satisfy a closed algebra with the usual definition (\ref{usualdefnambu}) \cite{Hasebe-2014-2}: 
\be
[O_{\mu_1}, O_{\mu_2}, \cdots, O_{\mu_{2k-1}}] = O_{[\mu_1}, O_{\mu_2}, \cdots, O_{\mu_{2k-1]}}. 
\ee
We need to introduce the chiral bracket defined by \cite{Jabbari2004,JabbariTorabian2005} 
\be
[O_{\mu_1}, O_{\mu_2}, \cdots, O_{\mu_{2k-1}}]_{\chi} \equiv [O_{\mu_1} O_{\mu_2} \cdots O_{\mu_{2k-1}}, \Gamma_{2k+1}], 
\label{chiralbracket}
\ee
where $\Gamma_{2k+1}$ denotes the chiral matrix of the $SO(2k)$ group.   
The use of the chiral bracket is favorable in the sense that it avoids problems inherent to the usual definition of the odd dimensional  bracket: 
 As briefly mentioned in Sec.\ref{sec:intro}, there exist difficulties in quantization and fundamental identity   in the  odd D Nambu bracket \cite{CurtrightZachos2003},  
while such difficulties do not arise in  even dimensions.  
Since the chiral bracket (\ref{chiralbracket}) is essentially $2k$-bracket, it can evade such problems \cite{Jabbari2004,JabbariTorabian2005}. 
It is easy to see that the quantized coordinates  $X_{\mu}$ (\ref{defcoors2k-1})  satisfy a closed algebra with the use of the chiral bracket: 
\be
[X_{\mu_1}, X_{\mu_2}, \cdots, X_{\mu_{2k-1}}]_{\chi}=-i^k C(k, I) ~\alpha^{2k-2} \epsilon_{\mu_1\mu_2 \cdots \mu_{2k}}X_{\mu_{2k}}  
\label{algebraoffuzzys2k-1nambu}
\ee
where 
\be
C(k, I)=\frac{(2k)!!(I+2k-2)!!}{2^{2k-1} I !!}. 
\ee
Eq.(\ref{algebraoffuzzys2k-1nambu}) represents the algebraic relation of $S_F^{2k-1}$ in quantum Nambu formalism, and it can also 
 be  derived  by ``dimensional reduction'' from the quantum Nambu algebra of $S_F^{2k}$ 
\cite{Hasebe-2014-1} 
\be
[X_{a_1}, X_{a_2}, \cdots, X_{a_{2k}}]=i^k C(k, I) ~\alpha^{2k-1} \epsilon_{a_1 a_2 \cdots a_{2k+1}}X_{a_{2k+1}}, 
\label{nambusf2k}   
\ee
with  identification $\Gamma_{2k+1} =\frac{1}{\alpha} X_{2k+1}$.  
In low dimensions, we have 
\begin{align}
&S_F^1~:~[X_{\mu_1}]_{\chi}=-i \epsilon_{\mu_1\mu_2} X_{\mu_2}, \nn\\
&S_F^3~:~[X_{\mu_1}, X_{\mu_2}, X_{\mu_3}]_{\chi}=\alpha^2 (I+2)~ \epsilon_{\mu_1\mu_2\mu_3\mu_4} X_{\mu_4}, \nn\\
&S_F^5~:~[X_{\mu_1}, X_{\mu_2}, X_{\mu_3}, X_{\mu_4}, X_{\mu_5}]_{\chi}=i\frac{3}{2}\alpha^4 (I+2)(I+4)~ \epsilon_{\mu_1\mu_2\mu_3\cdots \mu_6} X_{\mu_6}, \nn\\
&S_F^7~:~[X_{\mu_1}, X_{\mu_2}, X_{\mu_3}, X_{\mu_4}, \cdots, X_{\mu_7}]_{\chi}=-3 \alpha^6 (I+2)(I+4)(I+6)~ \epsilon_{\mu_1\mu_2\mu_3\cdots \mu_8} X_{\mu_8}. 
\end{align}
Square of radius of $S_F^{2k-1}$ is derived as  \cite{Ramgoolam2002,JabbariTorabian2005}
\be
\sum_{\mu=1}^{2k} X_{\mu}X_{\mu}=\frac{1}{2}\alpha^2(I+1)(I+2k-1)~\bs{1}_{2 d[0, \frac{I}{2}, \frac{1}{2}]_{SO(2k)}}. 
\ee

\subsection{Dimensional ladder}\label{sec:ladderofdim}

In the above, we introduced the chiral Nambu bracket  by ``embedding'' it in   one-dimension higher even-bracket.   Since the basic philosophy of non-commutative geometry is to realize quantum geometry by associated algebra, the algebraic embedding   implies a geometric  embedding of fuzzy odd-sphere in one-dimension higher non-commutative space.  
Here we discuss the geometric structure of  arbitrary odd dimensional fuzzy spheres  in the context of the Landau models, which naturally generalizes the preceding study of fuzzy three-sphere \cite{JabbariTorabian2005}. 

Recall that the lowest Landau level Hilbert space of $2k$D Landau model is spanned by the $SO(2k+1)$ fully symmetric representation $[0, \frac{I}{2}]_{SO(2k+1)}$ with dimension  \cite{Hasebe-2014-1}
\be
d[n=0, \frac{I}{2}]_{SO(2k+1)}=\prod_{l=1}^k \prod_{i=1}^l \frac{I+l+i-1}{l+i-1},  
\label{ladderofdimequ1}
\ee
which is  equal to the total dimension of the lowest Landau level of  the $(2k-1)$D Landau model (\ref{sumlambdadege}): 
\be
d[0, ~\frac{I}{2}]_{SO(2k+1)} =
\sum_{s=-\frac{I}{2}}^{\frac{I}{2}}  d[0,~ \frac{I}{2},~s]_{SO(2k)} . 
\label{sumoflambddegeso2k}
\ee
This implies a close relation between the $2k$ and $(2k-1)$D lowest Landau levels.  
Meanwhile from Eq.(\ref{n=0numberofstates}),  we obtain\footnote{In particular for the spinor representation, we have 
\be
d[0, \frac{1}{2}]_{SO(2k)}=2^{k-1}=d[0, \frac{1}{2},\frac{1}{2}]_{SO(2k-1)}. 
\label{rel2kand2k-1}
\ee
}    
\be
d[0, \frac{I}{2}, \pm \frac{I}{2}]_{SO(2k)} 
= d[0, \frac{I}{2}]_{SO(2k-1)}  
\label{rel2kand2k-gene}
\ee
indicating another relationship  between 
 $(2k-1)$ and $(2k-2)$D.  
Eq.(\ref{sumoflambddegeso2k}) and (\ref{rel2kand2k-gene}) suggest the dimensional ladder of the lowest Landau levels 
\be
2k\text{D LLL} ~~\overset{\text{Eq.}(\ref{sumoflambddegeso2k})}{\Longrightarrow} ~~(2k-1)\text{D LLL}~~ \overset{\text{Eq.}(\ref{rel2kand2k-gene})}{\Longrightarrow} ~~(2k-2)\text{D LLL}.
\ee
 
\begin{figure}[tbph]\center
\includegraphics*[width=120mm]{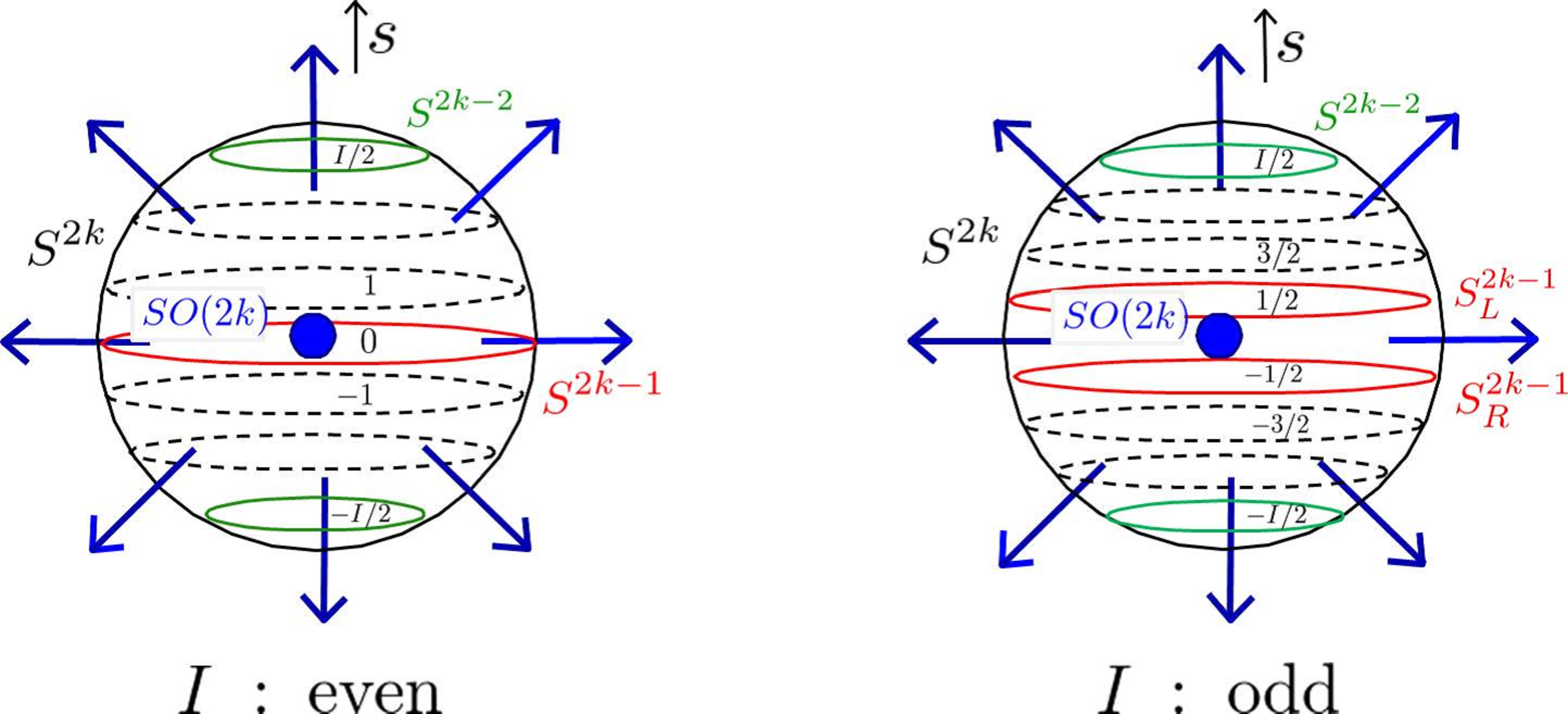}
\caption{The geometric picture of the fuzzy $2k$-sphere. 
The fuzzy $2k$-sphere is physically realized in the  $SO(2k)$ gauge monopole background \cite{Hasebe-2014-1}.   
 }
\label{EvenOddIfuzzyspheres}
\end{figure}

We first discuss a geometric interpretation of the relation (\ref{sumoflambddegeso2k}) between $2k$ and $(2k-1)$D.  In the $(2k-1)$D Landau model,  the lowest Landau level  eigenstates carry the internal chiral parameter $s$, which we will regard as a virtual  $x_{2k+1}$-axis.  Justification of this postulation is that,  for the lowest Landau level eigenstates, $s$ denotes the eigenvalues of the $(2k+1)$th coordinate of $S_F^{2k}$,  
\be
\frac{1}{2\alpha}X_{2k+1}=\frac{1}{2}\Psi^t \gamma_{2k+1} \frac{\partial}{\partial \Psi}=\frac{1}{2}\Psi^t_L \frac{\partial}{\partial \Psi_L} - \frac{1}{2}\Psi^t_R \frac{\partial}{\partial \Psi_R}.   
\ee
Geometrically, $s$ specifies the position of the hyper-latitude on the virtual $S^{2k}_F$
({Fig.~\ref{EvenOddIfuzzyspheres}}), and such  hyper-latitude  carries $d[0, \frac{I}{2}, s]_{SO(2k)}$ degrees of freedom
that represents the sub-band degeneracy of the lowest Landau level of the 
$(2k-1)$D Landau model. The sub-band degrees of freedom are stacked as
the hyper-latitudes along the $(2k+1)$th axis and in total they
constitute the $S_F^{2k}$. Each hyper-latitude corresponds to the lowest Landau level sub-band or the odd dimensional fuzzy sphere $S_F^{2k-1}$.\footnote{In a precise sense \cite{JabbariTorabian2005}, $(2k-1)$D fuzzy sphere  is realized as a superposition of two latitudes, $s=1/2$ and $s=-1/2$, nearest to the equator of fuzzy $S^{2k}$ for odd $I$ (the red circles in the right of Fig.~\ref{EvenOddIfuzzyspheres}).} 
%
As the magnitude of $s$ increases, $d[0, \frac{I}{2}, s]$ {(\ref{n=0numberofstates})} decreases. This coincides with our intuition that
the hyper-latitude $S^{2k-1}$ farther from the equator becomes smaller
to accommodate fewer degrees of freedom. 
The number of degrees of freedom on each hyper-latitude is given by 
\be
d[0, \frac{I}{2}, s]_{SO(2k)}~\sim~I^{\frac{1}{2}(k+2)(k-1)}  ~~~~(s<< I), 
\ee
and the number of the latitudes is $\sim I$.  Consequently, the total number of degrees of freedom of  $S^{2k}_F$ is roughly evaluated as 
\be
I^{\frac{1}{2}(k+2)(k-1)}\times I =I^{\frac{1}{2}k(k+1)}, 
\label{relationroughIlarge}
\ee
which essentially reproduces Eq.(\ref{sumoflambddegeso2k}) as\footnote{Considering that the number of the hyper-latitudes on $S_F^{2k}$ is $I+1= \text{deg}(S_F^2)$ and  $S^{2k-1}_F$ corresponds to each hyper-latitude, we can restate Eq.(\ref{relationroughIlarge}) as 
\be
\text{deg}(S_F^{2k-1})~\sim ~\text{deg}(S_F^{2k})/\deg(S_F^2).  
\ee
}
\be
\sum_{s=-\frac{I}{2}}^{\frac{I}{2}}  d[0,~ \frac{I}{2},~s]_{SO(2k)}~\sim~I^{\frac{1}{2}(k+2)(k-1)}\times I = I^{\frac{1}{2}k(k+1)}~\sim ~d[0, ~\frac{I}{2}]_{SO(2k+1)} .
\ee
Thus the dimensional ladder between the $2k$ and $(2k-1)$D Landau models (\ref{sumoflambddegeso2k}) can be  understood in the context of  non-commutative geometry.   Interestingly, we have $\it{not}$  a priori assumed the embedding of the odd dimensional Landau model in one-dimension higher space, but as  the analysis proceeds, the embedding structure has naturally emerged.   
The $SO(2k)$ gauge group in the winding number (\ref{mapfrom2k-1to2k}) now finds its origin too:  Since  $SO(2k)$ is the gauge group of the $2k$D Landau model  \cite{Hasebe-2014-1}, its appearance is accounted for by the embedding of the $(2k-1)$D Landau model in  the $2k$D Landau model.  

A simple geometric picture also explains the relation  between $(2k-1)$ and $(2k-2)$D (\ref{rel2kand2k-gene}). Even dimensional fuzzy sphere $S_F^{2k}$ is locally equivalent to  $S^{2k-2}_F$ fibration over $S^{2k}$: 
\be
S_F^{2k}~\sim~ S^{2k}\otimes S^{2k-2}_F, 
\ee
and then the number of the  degrees of freedom of the  latitude  around the north-pole (or the south-pole),  $d[0, \frac{I}{2}, \pm\frac{I}{2}]$, should coincide with that of  $S_F^{2k-2}$, 
$d[0, \frac{I}{2}]_{SO(2k-2)}$. 
Eq.(\ref{rel2kand2k-gene}) reflects such geometry (see Fig.\ref{EvenOddIfuzzyspheres}).  

 We thus have seen the following dimensional ladder: From the point of view of the group theory,  
\be
SO(2k+1)~\rightarrow~SO(2k)~\rightarrow~SO(2k-1), 
\ee
from the non-commutative geometry, 
\be
S_F^{2k}~\rightarrow~S_F^{2k-1}~\rightarrow~S_F^{2k-2},  
\ee
and from the lowest Landau level,  
\be
{2k} \text{D}~\text{LLL}~\rightarrow~{(2k-1)} \text{D}~\text{LLL}~\rightarrow~ ~{(2k-2)} \text{D}~\text{LLL}.
\ee
In particular for $k=2$, we have 
\begin{align}
SO(5) ~~~~&~\longrightarrow~SO(4)\simeq SU(2)_L\otimes SU(2)_R~~~~
\longrightarrow~~~~~~~ SU(2)_{L/R}, \nn\\
S_F^4~~~~~~&~\longrightarrow~~~~~~~~~~~~~~~~~~S_F^3~~~~~~~~~~~~~~~~~~~~\longrightarrow~~~~~~~~~~S_F^2, \nn\\
\text{4} \text{D} ~\text{LLL}~~~&~\longrightarrow~~~~~~~~~~~~~~~~\text{3} \text{D}~\text{LLL}~~~~~~~~~~~~~~~\longrightarrow~~~~~~~\text{2} \text{D} ~\text{LLL} . 
\label{hierark=2}
\end{align}
Eq.(\ref{hierark=2}) may remind the readers of the dimensional ladder of anomaly associated with  relativistic fermion on $S^4$ in   $SU(2)$ instanton background  \cite{Forte-1986, Forte-1987}. In these literatures,  instantons with gauge group of the fundamental representation ($I=1$) and non-trivial Potryagin index are adopted, while in the present work we deal with gauge  bundle  with unit Pontryagin index and gauge group of higher representation $(I\ge 1)$.\footnote{The instanton Chern number is generally given by the product of the Pontryagin index $N$ and  the gauge group representation part \cite{Jackiw-Rebbi-1977}:  
\be
c_2=N\cdot \frac{1}{6}I(I+1)(I+2). 
\ee
In Refs.\cite{Forte-1986, Forte-1987} $I=1$ and $N \ge 1$, while in the present paper $N=1$ and $I\ge 1$.  
}      
We will revisit the dimensional ladder in Sec.\ref{sec:atiyahpatodisinger}, but before that, we exploit the relativistic Landau models on odd dimensional spheres.  

\section{Relativistic Landau Models on $S^{2k-1}$}\label{landaudiracodd}

\subsection{Dirac-Landau models}

Here, we consider  Dirac-Landau models on $S^{2k-1}$. The $SO(2k)$ gamma matrices are taken as 
\be
\Gamma^{i}=\begin{pmatrix}
\gamma^i & 0 \\
0 & -\gamma^i 
\end{pmatrix}~(i=1,2,\cdots, 2k-1),~~~\Gamma^{2k}=\begin{pmatrix}
0 & 1_{2^{k-1}} \\
1_{2^{k-1}} & 0 
\end{pmatrix}, 
\ee
where $\gamma^i$ denote  $SO(2k-1)$ gamma matrices. 
The chirality matrix is given by 
\be
\Gamma^{2k+1}=\begin{pmatrix}
0 & i1_{2^{k-1}}  \\
-i 1_{2^{k-1}} & 0 
\end{pmatrix}. 
\ee
The Dirac-Landau operator on $S^{2k-1}$ is constructed as   
\be
-i\fsl{D}=-i\sum_{i=1}^{2k-1}\Gamma^{i}e_i^{~~\alpha}D_{\alpha}=
\begin{pmatrix}
-i\fsl{D}_{\text{W}} & 0\\
0 & i\fsl{D}_{\text{W}} 
\end{pmatrix},  \label{freedirachamils2k-1}
\ee
where $-i\fsl{D}_{\text{W}}$ denotes the Weyl-Landau operator (\ref{WeylLandauopedef}), and 
 the massive Dirac-Landau Hamiltonian is given by 
\be
\mathcal{H}=-i\fsl{D} +M\Gamma^{2k} =\begin{pmatrix}
-i\fsl{D}_{\text{W}} & M\\
M & i\fsl{D}_{\text{W}} 
\end{pmatrix}. 
\ee
 $\mathcal{H}$ is a Dirac-Landau Hamiltonian of the AIII class topological insulator  that respects the chiral symmetry, 
\be
\{\mathcal{H}, \Gamma^{2k+1}\}=0, 
\ee
and its spectrum is symmetric with respect to the zero-energy: 
\be
\pm \Lambda =\pm \sqrt{\lambda^2+M^2}, 
\ee
where $\lambda^2$ denote the eigenvalue of $(-i\fsl{D})^2$. 
From the results of  Sec.\ref{sec:weyllandau}, one can readily obtain the eigenvalues and the corresponding degeneracies of the Dirac-Landau operator.  The eigenvalues are the same as of the Weyl-Landau operator (\ref{deflambdasquares}): 
\be
(-i\fsl{D})^2= {\lambda_n^{+}(s)}^2,~~~{\lambda_n^{-}(s)}^2.    
\ee
For each of ${\lambda_n^{\pm }(s)}^2$, the degeneracy is equally given by 
\begin{align}
&s^2 >0~~~:~~2~d[n, \frac{I}{2}\pm \frac{1}{2}, s]_{SO(2k)},\nn\\
&s^2 =0~~~:~~d[n, \frac{I}{2}\pm \frac{1}{2}, 0]_{SO(2k)}. 
\end{align}
For the relativistic Landau models on $S^1$, we give a detailed analysis  in Appendix \ref{sec:reviewcircle}. 

\subsection{Weyl-Landau models}\label{sec:weyllandau}

 We analyze the Weyl-Landau operator on $S^{2k-1}$:  
\be
-i\fsl{D}_{\text{W}}=-i\sum_{i=1}^{2k-1}\gamma^{i}e_i^{~~\alpha}D_{\alpha}=\sum_{i=1}^{2k-1}\gamma^i (-i\partial_{\alpha}+{\omega}_{\alpha}+\mathcal{A}_{\alpha}), 
\label{WeylLandauopedef}
\ee
where $\omega_{\alpha}$ represent the $SO(2k-1)$ spin connection and   $\mathcal{A}_{\alpha}$ ($\alpha=x_1,x_2,\cdots, x_{2k-1}$) denote the $SO(2k-1)$ gauge field on $S^{2k-1}$ that corresponds to $A_{\mu}$ 
(\ref{so2k-1gaugemonopolefield}) in the Cartesian coordinate.  
Following Dolan's work \cite{Dolan2003},  we investigate the eigenvalue problem of the Weyl-Landau operator (\ref{WeylLandauopedef}).  On the coset $\mathcal{M}\simeq G/H$, the square of the Dirac (Weyl) operator is given by 
\be
(-i\fsl{D})^2=C(G)-C(H)+\frac{1}{8}\mathcal{R}_{\mathcal{M}}, 
\label{squareofDiracodd}
\ee
where $C$ denotes  quadratic Casimir of the corresponding group, and $\mathcal{R}_{\mathcal{M}}$ is the Ricci scalar  of $\mathcal{M}$. 
For the present case $S^{2k-1}\simeq SO(2k)/SO(2k-1)$, Eq.(\ref{squareofDiracodd}) becomes 
\be
(-i\fsl{D}_{\text{W}})^2=C_{SO(2k)}-C_{SO(2k-1)}+\frac{1}{8}{\mathcal{R}}_{S^{2k-1}}
\ee
where\footnote{For the fundamental representation $(I=1)$, the $SO(2k-1)$ Casimir is given by 
\be
\sum_{i <j=1}^{2k-1}{\sigma_{ij}}^2=\frac{1}{4}{(k-1)(2k-1)}=\frac{1}{8}{\mathcal{R}_{S^{2k-1}}}.
\ee
}  
\begin{align}
&C[n, J, s]_{SO(2k)}
=J^2(k-1)+J(2n+k^2-k)+n(n+2k-2)+s^2, \nn\\
&C[0, \frac{I}{2}]_{SO(2k-1)}
=\frac{I^2}{4}(k-1)+\frac{I}{2}(k-1)^2, \nn\\
&\mathcal{R}_{S^{2k-1}}=d(d-1)|_{d=2k-1}=2(k-1)(2k-1). 
\end{align}
By substituting these quantities to the formula (\ref{squareofDiracodd}) with 
\be
J=\frac{I}{2}\pm \frac{1}{2}, 
\ee
we have\footnote{
In the thermodynamic limit $I, R~\rightarrow~ \infty$ with fixed $s$ and $\omega \equiv {I}/R^2$, Eq.(\ref{deflambdasquares}) is reduced to 
\be
{\lambda^+_n( s)}^2~\simeq~{I (n+k-1)}, ~~~~~{\lambda^-_n( s)}^2~\simeq~{I n}. \label{thermolimitsn-}
\ee
} 
\begin{align}
{\lambda_n(s)}^2={\lambda_n(-s)}^2&=
\begin{pmatrix}
(n+k)(n+k-1)+I(n+k-1)+s^2 & 0 \\
0 & n(n+2k-3)+In +s^2
\end{pmatrix} \nn\\&\equiv 
\begin{pmatrix}
{\lambda_n^+(s)}^2 & 0 \\
0 & {\lambda^-_n(s)}^2 
\end{pmatrix}. 
\label{deflambdasquares}
\end{align}
For instance,\footnote{ For $SO(4)$, the eigenvalues (\ref{squareso4diraceigenval}) of $(-i\fsl{D}_{\text{W}})^2$ satisfy the special relation:  
\be
{\lambda_{n}^+(s)}^2={\lambda_{n+1}^-(s)}^2 = (n+2)(n+1)+I(n+1)+s^2~~~(n=0, 1, 2, \cdots), ~~~~~~ {\lambda_{0}^-(s)}^2=s^2, 
\label{patso4eigen}
\ee
with the degeneracies  
\be
d[n+\frac{I}{2}+\frac{1}{2}, s]_{SO(4)}=\frac{1}{4}(2n+I+3)^2-s^2 ~~~(n=0, 1, 2, \cdots), ~~~~~
d[\frac{I}{2}-\frac{1}{2}, s]_{SO(4)}=\frac{1}{4}(I+1)^2-s^2. 
\label{partso4dege}
\ee
These results coincide with Eq.(18) of Nair and Daemi \cite{Nair&Daemi2004} by  the replacement:  
\be
(J, q, \mu, T, S) ~\rightarrow~(\frac{I}{2}\pm \frac{1}{2}, n, \frac{I}{2}\pm \frac{1}{2}+\frac{\lambda}{2},  \frac{I}{2}, \frac{1}{2}). 
\ee
}
\begin{align}
&SO(4)~(k=2):~
\begin{pmatrix}
{\lambda_n^+(s)}^2 & 0 \\
0 & {\lambda^-_n(s)}^2 
\end{pmatrix}
=
\begin{pmatrix}
(n+2)(n+1)+I(n+1)+s^2 & 0 \\
0 & n(n+1) +In+s^2
\end{pmatrix}, \label{squareso4diraceigenval}\nn\\
&SO(6)~(k=3):~
\begin{pmatrix}
{\lambda_n^+(s)}^2 & 0 \\
0 & {\lambda^-_n(s)}^2 
\end{pmatrix}
=
\begin{pmatrix}
(n+3)(n+2)+I(n+2)+s^2 & 0 \\
0 & n(n+3) + In+s^2
\end{pmatrix}, \nn\\
&SO(8)~(k=4):~
\begin{pmatrix}
{\lambda_n^+(s)}^2 & 0 \\
0 & {\lambda^-_n(s)}^2 
\end{pmatrix}
=
\begin{pmatrix}
(n+4)(n+3)+I(n+3)+s^2 & 0 \\
0 & n(n+5) +In+s^2
\end{pmatrix}. 
\end{align}
 For the eigenvalues ${\lambda_n^{\pm}(s)}^2$ (\ref{deflambdasquares}), the corresponding degeneracies are  given by 
\be
2 (1-\delta_{s, 0})~d[n, \frac{I}{2}\pm \frac{1}{2}, s]_{SO(2k)}+\delta_{s ,0} ~d[n, \frac{I}{2}\pm \frac{1}{2}, 0]_{SO(2k)}, 
\label{lambdasqdege}
\ee
with $d_{SO(2k)}$ (\ref{degelandaunlambda}). The coefficient $2$ of the first term comes from the contribution of $+s$ and $-s$. 
 
Depending on the parity of $I$, the system exhibits different behaviors. 
For odd $I$,  $s$ can take $s=0$, and  the representations, $[n, \frac{I}{2}+\frac{1}{2}, 0]_{SO(2k)}$ and $[n,\frac{I}{2}-\frac{1}{2}, 0]_{SO(2k)}$, respectively yield the  eigenvalues
\begin{align}
&{\lambda^+_{n}(0)}^2= {(n+k+I)(n+k-1)}, \nn\\
&{\lambda^-_{n}(0)}^2={n(n+2k-3+I)}. 
\end{align}
Thus zero-modes are realized as $\lambda_{n=0}^-(0)=0$ for the representation  $[0, \frac{I}{2}-\frac{1}{2}, 0]_{SO(2k)}$.
Meanwhile for even $I$,  
the representations, $[n, \frac{I}{2}+\frac{1}{2}, \pm\frac{1}{2}]_{SO(2k)}$ and $[n, \frac{I}{2}-\frac{1}{2}, \pm\frac{1}{2}]_{SO(2k)}$, respectively yield the eigenvalues 
\begin{align}
&{\lambda^+_{n}(s=\pm{1}/{2})}^2= {(n+k-\frac{1}{2})^2+  I(n+k-1)}, \nn\\
&{\lambda^-_{n}(s=\pm1/2)}^2= {n(n+2k-3+I) +\frac{1}{4}}. 
\label{evenIspectrumdirac}
\end{align}
The corresponding degeneracies are summarized in Appendix \ref{sec:weyllandaudet}. Notice that the minimum value is not zero, but ${\lambda^-_{n=0}(s=\pm1/2)}^2= \frac{1}{4}.$ 

In the limit $I\rightarrow 0$, the former representation  $[n, \frac{I}{2}+\frac{1}{2},  \pm\frac{1}{2}]_{SO(2k)}|_{I=0} = [{n,\frac{1}{2}, \pm\frac{1}{2}}]_{SO(2k)}$ 
gives the spectrum of the free spinor particle on $S^{2k-1}$:\footnote{For each of the eigenvalues (\ref{ilimitzerofreedirac}), $+(n+k-\frac{1}{2})$ and $-(n+k-\frac{1}{2})$, the corresponding degeneracy is equally given by  
\be
d[{n, {1}/{2}}, {1}/{2}]_{SO(2k)}=2^{k-1}
\begin{pmatrix}
n+2k-2\\
n
\end{pmatrix} ~~~(n=0, 1, 2, \cdots)
\ee
which reproduces the known results of the free Dirac operator on $S^{2k-1}$ (see Ref.\cite{CamporesiHiguchi1996} for instance). 
  }
\be
\pm \lambda^+_{n}(\pm 1/2) ~\overset{I\rightarrow 0}{\longrightarrow}~-i\fsl{\nabla}=\pm (n+k-\frac{1}{2}), 
\label{ilimitzerofreedirac}
\ee
while 
the latter representation  $[n,  \frac{I}{2}-\frac{1}{2}, \pm\frac{1}{2}]_{SO(2k)}$  is not well defined, and then $\pm \lambda^-_{n}(s=\pm 1/2)$ do not have  their counterparts in the spectrum of the free Dirac operator. 

\subsection{Total degeneracy of $0$th Landau level  revisited}

For $n=0$, ${\lambda_{n}^-(s)}^2$ (\ref{deflambdasquares}) takes a very simple form:  
\be
{ \lambda_{n=0}^-(s)}^2=s^2, 
\ee
with  degeneracy
\be
 2 (1-\delta_{s, 0})~d[0, \frac{I}{2}-\frac{1}{2}, s]_{SO(2k)} +\delta_{s, 0} ~d[0, \frac{I}{2}-\frac{1}{2}, 0]_{SO(2k)} 
\ee
where 
\be
d[0, \frac{I}{2}-\frac{1}{2}, s]_{SO(2k)}=\prod_{j=1}^{k-1} \frac{(I+2j-1)^2 -4s^2}{(2j)^2}\prod_{l=1}^{k-2}\prod_{j=1}^{k-l-1}\frac{I+2l+j-1}{2l+j}.
\ee
Then  we have  
\begin{align}
&\text{total dimension of the $0$th Landau level}\nn\\
&= 2\sum_{s>0}^{\frac{I}{2}-\frac{1}{2}}d[0, \frac{I}{2}-\frac{1}{2}, s]_{SO(2k)} ~ \biggl(+ ~d[0, \frac{I}{2}-\frac{1}{2}, 0]_{SO(2k)} \biggr)
=  \sum_{s=-\frac{I}{2}+\frac{1}{2}}^{\frac{I}{2}-\frac{1}{2}} d[0, \frac{I}{2}-\frac{1}{2}, s]_{SO(2k)}\nn\\
&=  \nu_k(I) , 
\label{totaldimzerollnu}
\end{align}
where   Eq.(\ref{indextheoremvariant}) was used in the last equation.  Zero-mode contribution $(~~)$ on the second equation  exists when  $I$ is odd.  
We will explain a differential topology origin behind this relation.   
First notice, since  $d[0, \frac{I}{2}-\frac{1}{2}]_{SO(2k+1)}$ denotes the 0-mode degeneracy of the $2k$D relativistic Landau model,  $i.e.$,  the index of the Dirac-Landau operator on $S^{2k}$,   the dimensional ladder between  $2k$ and $(2k-1)$D (\ref{sumoflambddegeso2k}) can be rewritten as 
\be
\sum_{s=-\frac{I}{2}+\frac{1}{2}}^{\frac{I}{2}-\frac{1}{2}} d[0, \frac{I}{2}-\frac{1}{2}, s]_{SO(2k)} = \text{Ind}[-i\fsl{D}_{SO(2k+1)}]. 
\label{fromfuzzysphererel}
\ee
Second, we find that the Atiyah-Singer index theorem tells\footnote{On general spin manifold without boundary, the Atiyah-Singer theorem is given by 
\be
\text{Ind}(-i\fsl{D}) =\int_{M}\hat{A}(R)\cdot ch(F), 
\ee
where $ch(F)$ denotes the Chern number density and  $\hat{A}(R)$ represents the Dirac-genus given by the polynomials of the Pontryagin class: 
\be
\hat{A}(R) =1-\frac{1}{24}p_1(R) +\frac{1}{5760} (7{p_1(R)}^2-4p_2(R)) +\cdots. 
\ee
 For spheres, the Pontryagin class is identically zero, and so the Dirac-genus becomes  trivial.  Consequently for $S^{2k}$, the index of the Dirac operator is determined solely  by the gauge bundle topology: 
\be
\text{Ind}(-i\fsl{D}) =\int_{S^{2k}} ch(F)=c_k. 
\ee
  } 
\be
\text{Ind}[-i\fsl{D}_{SO(2k+1)}]=c_k(I), 
\label{indextheoremndc}
\ee
where $c_k$ denotes the $k$th Chern number of the $SO(2k)$ monopole charge. For the $2k$D Landau models, Eq.(\ref{indextheoremndc}) can be verified explicitly \cite{Hasebe-2014-1}. 
Third, from the fibre-bundle theory,  the non-trivial gauge topology on $2k$-sphere is evaluated by the winding number (\ref{windingnumberexpli}) from the transition function on the  equator $S^{2k-1}$: 
\be
c_{k}(I)=\nu_{k}(I). 
 \label{candwindingrelation}
\ee
See Appendix \ref{append:wigingchern} about a  general proof of Eq.(\ref{candwindingrelation}). 
In the present case,  we can explicitly verify Eq.(\ref{candwindingrelation}) as follows. Using the formula of  $d[0, \frac{I}{2}, s]_{SO(2k-2)}$   (\ref{n=0numberofstates}), we perform the the summation of the winding number (\ref{fomulaforwindingexp})  to have 
\be
\nu_k(I)= \frac{1}{2^{k-1}} \frac{(I+2k-4)!!}{(2k-3)!! I!!} \sum_{s=-\frac{I}{2}}^{\frac{I}{2}}(2s)^2 d[0,\frac{I}{2}, s]_{SO(2k-2)} = \prod_{l=1}^k \prod_{i=1}^{l}\frac{I+l+i-2}{l+i-2}=d[0, \frac{I-1}{2}]_{SO(2k+1)}, 
\ee
where  $d[0, \frac{I}{2}]_{SO(2k+1)}$ is defined by Eq.(\ref{ladderofdimequ1}).  In Ref.\cite{Hasebe-2014-1}, we  showed  $d[0, \frac{I-1}{2}]_{SO(2k+1)}=c_{k}(I)$,  and then 
\be
c_k(I)=d[0, \frac{I-1}{2}]_{SO(2k+1)} =\nu_{k}(I). 
\label{fromfibrebund}
\ee
Finally, from Eq.(\ref{fromfuzzysphererel}), (\ref{indextheoremndc}) and (\ref{candwindingrelation}), we obtain the sequence 
\be
\!\!\!\!\!\!\text{total dim. of  odd D LLL} \overset{\text{dim. ladder (\ref{fromfuzzysphererel})}}{\Longrightarrow} ~\text{Ind}[-i\fsl{D}]_{\text{even D}} ~\overset{\text{AS index theorem (\ref{indextheoremndc})}}{\Longrightarrow} ~c~\overset{\text{fibre-bundle theory (\ref{candwindingrelation})}}{\Longrightarrow}~ \nu ,  
\label{primaryseq}
\ee
which verifies Eq.(\ref{totaldimzerollnu}) after all. 
However, there is one speck in the above discussions.    
As for the dimensional ladder (\ref{fromfuzzysphererel}),  we utilized  specific properties of the Landau models, while the other processes of the sequence depend on general arguments of differential topology,  such as Atiyah-Singer index theorem and fibre-bundle theory. We will reconsider  the dimensional ladder  and show that the dimensional ladder can also be understood from a differential  topology point of view. 
In Sec.\ref{sec:atiyahpatodisinger}, the total dimension of the 0th Landau level is shown to be equal to the spectral flow: 
\be
\sum_{s} d[0, \frac{I}{2}-\frac{1}{2}, s]_{SO(2k)} =- ~\text{spectral flow} .   
\label{spectraldimso2k-1}
\ee
With Eq.(\ref{spectraldimso2k-1}) and the spectral flow argument of Atiyah-Patodi-Singer   \cite{Atiyah-Patodi-Singer-1975-1, Atiyah-Patodi-Singer-1975-2, Atiyah-Patodi-Singer-1976, AlvarezGaume-Pietra-Moore-1985}  
\be
- ~\text{spectral flow}  =\text{Ind}(-i\fsl{D}), 
\label{spectralflowAPS}
\ee
 every process of the sequence (\ref{primaryseq}) can be restated in the language of  differential topology: 
\be
\!\!\!\!\!\!\text{total dim. of  odd D LLL} \overset{\text{APS spectral flow  (\ref{spectralflowAPS})}}{\Longrightarrow} ~\text{Ind}[-i\fsl{D}]_{\text{even D}} ~\overset{\text{AS index theorem (\ref{indextheoremndc})}}{\Longrightarrow} ~c~\overset{\text{fibre-bundle theory (\ref{candwindingrelation})}}{\Longrightarrow}~ \nu.   
\ee

\section{Ladder of Dimensions: Spectral Flow  of Atiyah-Patodi-Singer}\label{sec:atiyahpatodisinger}

Let us discuss relations between the spectral flow argument of  Atiyah-Patodi-Singer  and the dimensional ladder between the $2k$ and  $(2k-1)$D Landau models: 
\be
d[0, ~\frac{I}{2}]_{SO(2k+1)} =\sum_{s=-\frac{I}{2}}^{\frac{I}{2}}  d[0,~ \frac{I}{2},~s]_{SO(2k)} . 
\label{sumoflambddegeso2k-2}
\ee
 We demonstrate how Eq.(\ref{sumoflambddegeso2k-2}) works in   1D and 2D relativistic Landau models and clarify relations to the spectral flow argument.   
 For this purpose, we apply  the  Atiyah's discussion \cite{Atiyah-1985} originally developed on  torus $S^1\otimes S^1$ is applied to fuzzy two-sphere.\footnote{Also see  Refs.\cite{AsoreyEstevePacheco1983, AguadoAsoreyEsteve2001} for relations to quantum anomaly. }  
 In this section, we use  the terminologies ``lowest Landau level'' and ``0th Landau level'' interchangeably.

\subsection{2D and 1D relativistic Landau models}

\begin{figure}[tbph]\center
\includegraphics*[width=130mm]{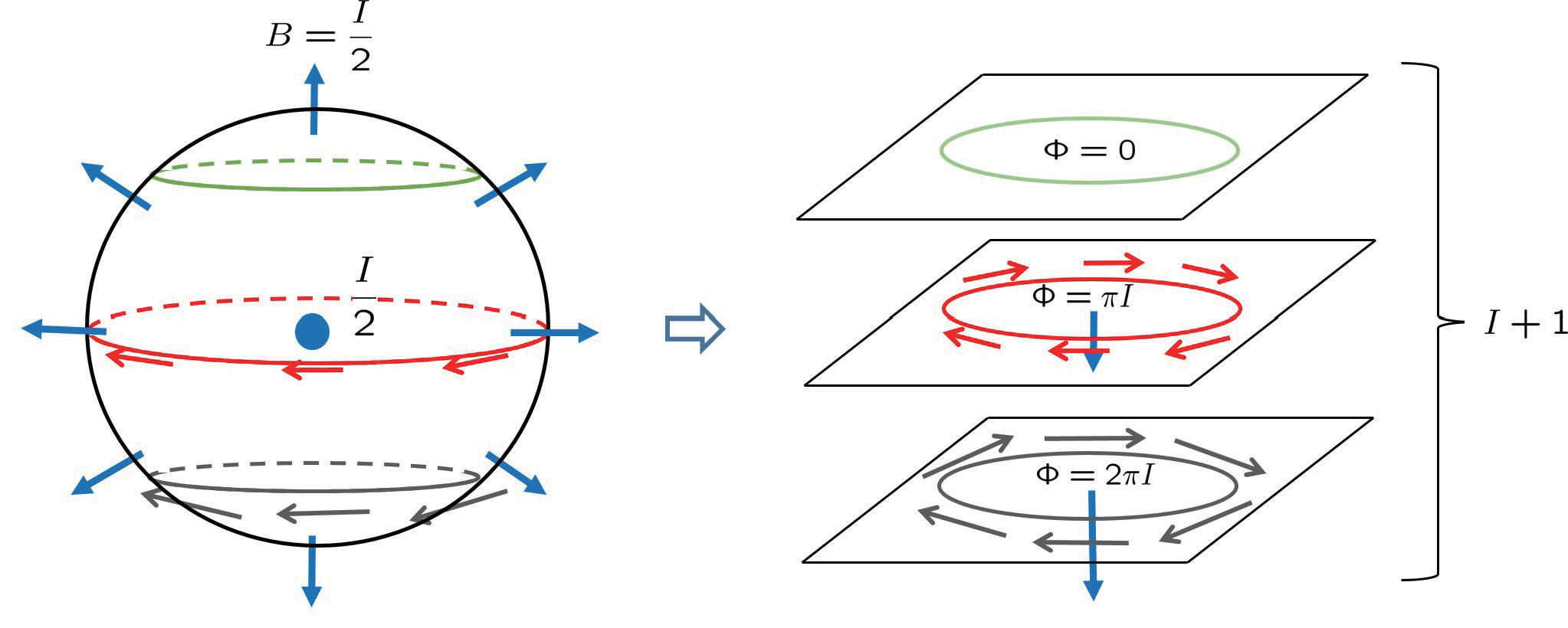}
\caption{1D Landau models as  latitudes of the 2D Landau model.  
  }
\label{From2Dto1D.fig}
\end{figure}

The massive Dirac-Landau operator on two-sphere is given by 
\be
-i\fsl{D}_{S^2}+M\sigma_z = -i\sigma_x (\partial_{\theta}-\frac{1}{2}\tan\frac{\theta}{2}) -i\frac{1}{\sin\theta} \sigma_y (\partial_{\phi} -iA_{\phi}) +M\sigma_z, 
\label{2DDiracopLandau}
\ee 
where $\theta$ and $\phi$ represent the standard polar coordinates, $0\le \theta < \pi$, $0\le \phi < 2\pi$, and 
$A_{\phi}$ denotes the monopole gauge field
\be
A=A_{\phi}d\phi=\frac{I}{2}(1-\cos\theta) d\phi. 
\ee
We omit the spin connection part to deduce a massive 1D Dirac-Landau operator from Eq.(\ref{2DDiracopLandau}):   
\be
-i\fsl{D}_{S^1}+M\sigma_z=-i\sigma_y (\partial_{\phi} - iA_{\phi}  )+M\sigma_z\nn\\
= 
\begin{pmatrix}
M & -\partial_{\phi}+iA_{\phi} \\
\partial_{\phi}-iA_{\phi} & -M 
\end{pmatrix}  
\label{chiral1dtoopdirac}
\ee
that  respects the chiral symmetry: 
\be
\{\sigma_x, -i\fsl{D}_{S^1}+M\sigma_z\}=0.   
\ee
($-i\fsl{D}_{S^1}+M\sigma_z$ denotes a 1D chiral topological insulator Hamiltonian.\footnote{
Without the gauge field, Eq.(\ref{chiral1dtoopdirac}) is reduced to  the Jackiw-Rebbi Hamiltonian used for the low energy description of  polyacetylene \cite{Jackiw-Rebbi-1976}.})    
The 1D  Dirac-Landau operator $-i\fsl{D}_{S^1}$  
\be
-i\fsl{D}_{S^1}=-i\sigma_y D_{S^1}=\begin{pmatrix}
0 &  -D_{S^1}  \\
 D_{S^1}   & 0 
\end{pmatrix}  
\label{twosum1ddirac}
\ee
consists of two copies of the 1D Weyl-Landau operator 
\be
-iD_{S^1}=-i\partial_{\phi} - A_{\phi}^{(\theta)},  
\label{1DDiracopLandau}
\ee
with
\be
A^{(\theta)}_{\phi}=\frac{I}{2}(1-\cos\theta). 
\ee
Notice that in the 1D Weyl-Landau operator (\ref{1DDiracopLandau}), $\theta$ is no longer a coordinate but denotes an external parameter to tune the magnitude of the gauge field.  
While magnetic field is constant on the surface of the two-sphere,   Eq.(\ref{1DDiracopLandau}) describes the 1D Weyl operator with the magnetic flux at the center of $S^1$ (Fig.\ref{From2Dto1D.fig}). 
The magnitude of the magnetic flux is given by  
\be
\Phi^{(\theta)}=\oint d\phi A^{(\theta)}_{\phi}=2\pi A^{(\theta)}_{\phi} =\pi I (1-\cos\theta) , 
\label{phifluxmag}
\ee
which monotonically increases from $0$ to $2\pi I$ as the parameter $\theta$ changes from $\theta=0$ to $\pi$. 

\subsubsection{Atiyah-Singer index theorem}

There are not a few literatures about the  non-relativistic/relativistic Landau problem on $S^2$ (see Ref.\cite{Hasebe-2015} for instance).  
In the non-relativistic case,  the lowest Landau level degeneracy is given by  
\be
d_{\text{LLL}}^{2\text{D}}(\frac{I}{2})=d[0, \frac{I}{2}]_{SO(3)}=I+1, 
\label{nonrelas2landau}
\ee
and the lowest Landau level basis states are the eigenstates of $X_z=\frac{2}{I}L_z$  with the eigenvalues   
\be
X_z=1 -\frac{2}{I}m ~~~~~(m=0,1,2,\cdots, I).
\ee
In other words, the lowest Landau level basis states correspond to the latitudes with the  azimuthal angles $\theta_m$: 
\be
\cos\theta_m=1-\frac{2}{I}m. 
\label{quantizatinthetas2}
\ee
The zero-mode degeneracy of the Dirac-Landau operator is obtained from the lowest Landau level degeneracy (\ref{nonrelas2landau}) with the replacement  $ \frac{I}{2}~\rightarrow ~\frac{I}{2}-\frac{1}{2}$, and the index of the Dirac-Landau operator is given by    
\be
\text{Ind}(-i\fsl{D}_{S^2})=d_{\text{LLL}}^{2\text{D}}(\frac{I}{2}-\frac{1}{2})=I. 
\label{index2ddirac}
\ee
Meanwhile, the 1st Chern number of the monopole bundle  is obtained as  
\be
c_1(g)=\frac{1}{2\pi}\int_{S^2} F= I. 
\ee
Their equality manifests the Atiyah-Singer index theorem: 
\be
\text{Ind}(-i\fsl{D}_{S^2})=c_1. 
\ee

\subsubsection{Spectral flow argument}

\begin{figure}[tbph]\center
\includegraphics*[width=80mm]{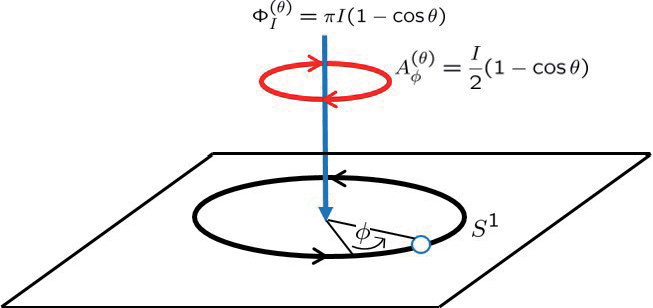}
\caption{The flux insertion reduces the particle angular momentum  due to the associated counter-propagating gauge field. }
\label{1DFluxIns.fig}
\end{figure}

Next let us consider the 1D Weyl-Landau model. 
With the magnetic flux $\Phi^{(\theta)}$ (\ref{phifluxmag}), the 1D Weyl-Landau operator is  represented as 
\be
-iD_{S^1}=-i(\partial_{\phi} -i\frac{1}{2\pi}\Phi^{(\theta)})=L_{\phi},  
\label{1Dweylphi}
\ee
which is simply the covariant derivative in $\phi$-direction or the covariant angular momentum, $L_{\phi}$.  
Intuitively, the flux insertion brings the counter-propagating angular momentum along $\phi$ direction (Fig.\ref{1DFluxIns.fig}).  
Since  $\theta$ is a continuum variable, the magnitude of the flux  $\Phi^{(\theta)}=\pi I(1-\cos\theta)$, 
 is not generally quantized unlike the monopole charge. 
 However, with respect to the latitudes 
{(\ref{quantizatinthetas2})},  the magnetic flux {(\ref{phifluxmag})} is quantized as a
multiple of unit flux:
\be
\Phi^{(\theta_m)}=2\pi  m ~~~~(m=0, 1, 2, \cdots, I). 
\label{quantizedflux}
\ee
Thus by embedding, the magnetic charges of the odd D Landau models  become quantized.

We solve the eigenvalue problem of the Weyl-Landau operator, 
\be
-iD_{S^1} \psi_n(\phi)=\lambda_n^{(\theta)} \psi_n(\phi). 
\ee
Due to the periodic boundary condition, the eigenfunctions are obtained as  
\be
\psi_n(\phi)=\frac{1}{\sqrt{2\pi}}e^{in\phi}~~~~~~~(n=0, \pm 1, \pm 2, \cdots),
\ee
with the eigenvalues  
\be
\lambda_n^{(\theta)}=n-\frac{1}{2\pi}\Phi^{(\theta)}=n-\frac{I}{2}(1-\cos\theta). 
\label{spectralfloweq}
\ee
The 1D Weyl-Landau model on the latitude {(\ref{quantizatinthetas2}) takes the 
energy eigenvalues  
\be
\lambda_n^{(\theta_m)} =n-m.  
\label{nandm1d}
\ee

When $\theta$ changes from $0$ to $\pi$ (or   
the magnetic flux $\Phi^{(\theta)}$  increases  adiabatically from $0$ to $2\pi I$),   
 the spectrum evolves  from $n$ to $(n-I)$ according to Eq.(\ref{spectralfloweq})  (Fig.\ref{SpectralFlow.fig}). 
\begin{figure}[tbph]\center
\includegraphics*[width=160mm]{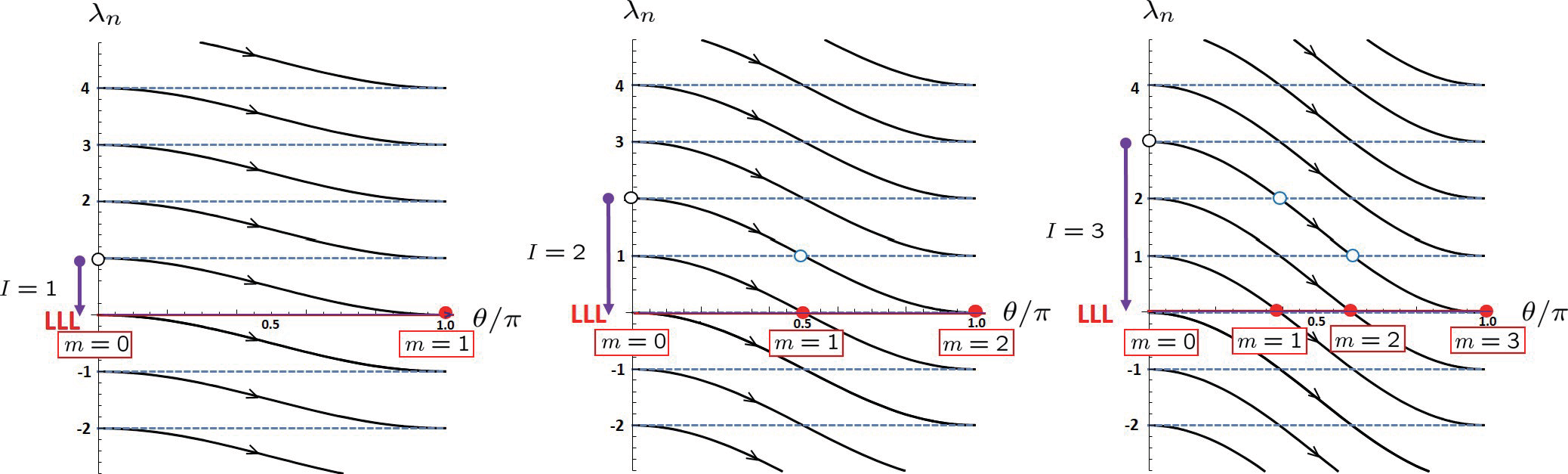}
\caption{Evolving eigenvalues (\ref{spectralfloweq}) during the process from $\theta=0$ to $\theta=\pi$ with respect  to  $I/2={1}/{2}, 1, {3}/{2}$ from left to right. 
 The eigenvalues  cross the zero-energy  (the red horizontal line) $I$ times at the points indicated by the red filled circles, the  number of which  is the definition of the spectral flow. 
Since each Landau level reduces its energy by $I$ during the process, 
the spectral flow is equal to the descent of  energy level (the length of the purple down-arrow).    
 }
\label{SpectralFlow.fig}
\vspace{3mm}
\end{figure}
The flux insertion brings the counter-propagating gauge field during the process, and  the particle on $S^1$ reduces its angular momentum and travels down from the $n$th circle to the  $(n-I)$th inner circle. 
 Since the energy transition of 1D Landau model is equal  to the number of the points at which the evolving eigenvalues cross  the zero-energy horizontal line (Fig.\ref{SpectralFlow.fig}),  we find that $I$ is equal to (the minus of) the spectral flow:\footnote{The sign of the spectral flow is taken so that each positive/negative eigenvalue that becomes  negative/positive  contributes -1/+1 to the spectral flow counting.}      
\be
- ~\text{spectral flow} = I. 
\ee
  
Recall that $I$ was the index of the 2D Dirac-Landau operator (\ref{index2ddirac}). 
Then we have 
\be
\text{Ind}(-i\fsl{D})=- ~\text{spectral flow}, 
\label{simpleaps}
\ee
which manifests the spectral flow argument of Atiyah-Patodi-Singer. 

\subsection{Dimensional ladder of non-commutative geometry}

In the above, we have seen  the validity of the  spectral flow argument in the context of the low dimensional Landau models, while in the following we will show that  the spectral flow argument generally guarantees the dimensional ladder  of  the Landau models (\ref{sumoflambddegeso2k-2}).  
The spectral flow indicates the number of points at which the evolving eigenstates  cross the zero-energy. In view of the 1D Landau models,  each crossing point  corresponds to the lowest Landau level of the Landau model with different magnetic flux.  Concretely, the $m$th crossing point corresponds to the lowest Landau level of the  Weyl-Landau model (\ref{1Dweylphi}) with flux $\Phi^{(\theta_m)}=2\pi m $ (Fig.\ref{SpectralFlow.fig}).  Since the spectral flow indicates the number of the crossing points, the  spectral flow is equal to  the  total number of the lowest Landau level states of $I$ 1D Weyl-Landau models with 
 $\Phi^{(\theta_m)}=2\pi m $  $(m=1, 2, \cdots, I)$: 
\be
- ~\text{spectral flow} = \sum_{m=1}^{I}d_{\text{LLL}}^{1\text{D}}( m) ,  
\label{etaandLLLdege}
\ee
where $d_{\text{LLL}}^{1\text{D}}(m)$ denotes the lowest Landau level degeneracy of the Weyl-Landau model with   $\Phi^{(\theta_m)}$,  
 and in the present case, obviously  $d_{\text{LLL}}^{1\text{D}}(m)=1$. 
  Besides,  the sum of the lowest Landau level states of the $I$ Weyl-Landau models can be given by the sum of the states in different energy levels of a $\it{single}$ Weyl-Landau  model.   
Notice that the  lowest Landau level of the Weyl-Landau model for $m$ and   
$s(= m-\frac{I}{2}-\frac{1}{2})$th  level of the Weyl-Landau model with  $\Phi^{(\theta_{m={I}/{2} -{1}/{2}})}=\pi (I-1)$  are on the same evolving eigenvalue curve (see Fig.\ref{thetapidiac.fig}):  
\be
d_{\text{LLL}}^{1\text{D}}( m) =d^{1\text{D}}_{s=m-\frac{I}{2}-\frac{1}{2}}( \frac{I}{2}-\frac{1}{2}), 
\ee
where $d^{1\text{D}}_{s}( \frac{I}{2}-\frac{1}{2}) $ denotes the $s$th level degeneracy of the Weyl-Landau model with flux $\Phi/(2\pi)=\frac{I}{2}-\frac{1}{2}$. 
\begin{figure}[tbph]\center
\includegraphics*[width=65mm]{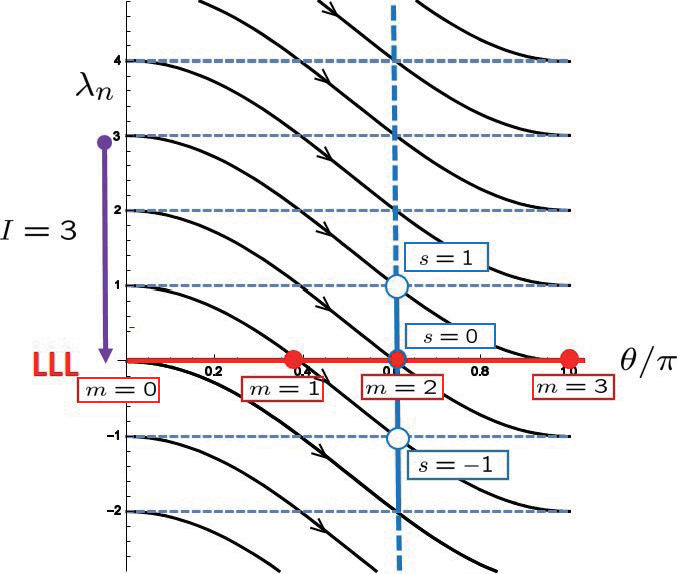}
\caption{Evolving eigenvalues of the 1D Weyl-Landau operator for ${I}={3}$.  
The intersections (blue circles) between the blue  vertical line and the energy curves represent the 
  $s(=-1, 0, 1)$th levels of the Weyl-Landau operator with $\Phi^{(\theta_{m=\frac{I}{2}-\frac{1}{2}})}=\pi (I-1)$. 
The  intersections (red filled circles)  between the zero-energy horizontal line and the energy curves  indicate the lowest Landau levels of the Weyl-Landau models with $\Phi^{(\theta_{m=1, 2, \cdots , I})}= 2\pi, 4\pi,  \cdots, 2\pi I$.  
The blue circle and the red filled one  on  same eigenvalue curve are related by $s=m-\frac{I}{2}-\frac{1}{2}$. }
\label{thetapidiac.fig}
\end{figure}
Hence, the spectral flow (\ref{etaandLLLdege}) can be  expressed by the sum of the  level degeneracies of the single Weyl-Landau model with $\Phi/(2\pi)=\frac{I}{2}-\frac{1}{2}$,  
\be
- ~\text{spectral flow}=  \sum_{s=-\frac{I}{2}+\frac{1}{2}}^{\frac{I}{2}-\frac{1}{2}} d^{1\text{D}}_{s}(\frac{I}{2}-\frac{1}{2}). 
\label{etaand1ddege}
\ee
This relation is naturally  generalized to higher dimensions as  
\be
- ~\text{spectral flow} 
 = \sum_{s=-\frac{I}{2}+\frac{1}{2}}^{\frac{I}{2}-\frac{1}{2}} d^{(2k-1)\text{D}}_{s}(\frac{I}{2}-\frac{1}{2})=\sum_{s=-\frac{I}{2}+\frac{1}{2}}^{\frac{I}{2}-\frac{1}{2}} d[0, \frac{I}{2}-\frac{1}{2}, s]_{SO(2k)},  
\ee
which is the sought formula (\ref{spectraldimso2k-1}) that relates the odd D Landau level degeneracies  and the spectral flow.  
Besides, the spectral flow argument  tells 
\be 
\text{Ind}(-i\fsl{D})=- ~\text{spectral flow}, 
\ee
and we eventually have 
\be
d[0, \frac{I}{2}-\frac{1}{2}]_{SO(2k+1)}  =\text{Ind}(-i\fsl{D})=- ~\text{spectral flow}   =\sum_{s=-\frac{I}{2}+\frac{1}{2}} ^{\frac{I}{2}-\frac{1}{2}}d[0, \frac{I}{2}-\frac{1}{2},   s]_{SO(2k)}.   
\label{sequence2k-1to2k}
\ee
This is the very formula of the dimensional ladder (\ref{sumoflambddegeso2k-2}). 
As discussed above,  the spectral flow argument of Atiyah-Patodi-Singer guarantees   the validity of the dimensional ladder of the Landau models.   
In other words, the dimensional ladder of non-commutative geometry can thus find its origin  in the  differential topology.     

We recapitulate the above discussions  with the 1D and 2D Landau models. 
From the two-sphere point of view,   $\theta$  is the azimuthal angle, and  the change of $\theta$ from $0$ to $\pi$ corresponds to a sweep on $S^2$ from north pole to  the south pole transversing $I$ latitudes, which is the dimension of  the 2D relativistic lowest Landau level, $d_{\text{LLL}}^{2\text{D}}(\frac{I}{2}-\frac{1}{2})$.  
During the sweep, the magnetic flux  changes from $0$ to $2\pi I$, and then each energy level of the 1D Weyl-Landau model decreases by $I$ that is the spectral flow. The spectral flow is  equal to the total dimension of the corresponding 1D Landau levels.     
We can thus restate Eq.(\ref{sequence2k-1to2k}) as    
\be
\text{2D LLL dim.} 
\Rightarrow \text{sweep on (fuzzy) two-sphere}  \Rightarrow \text{spectral flow} \Rightarrow  \text{total dim. of  1D LLs}. 
\label{etaand2ddege}
\ee
Fig.\ref{2D1D.fig} shows (\ref{etaand2ddege}) pictorially. 
\begin{figure}[tbph]\center
\includegraphics*[width=160mm]{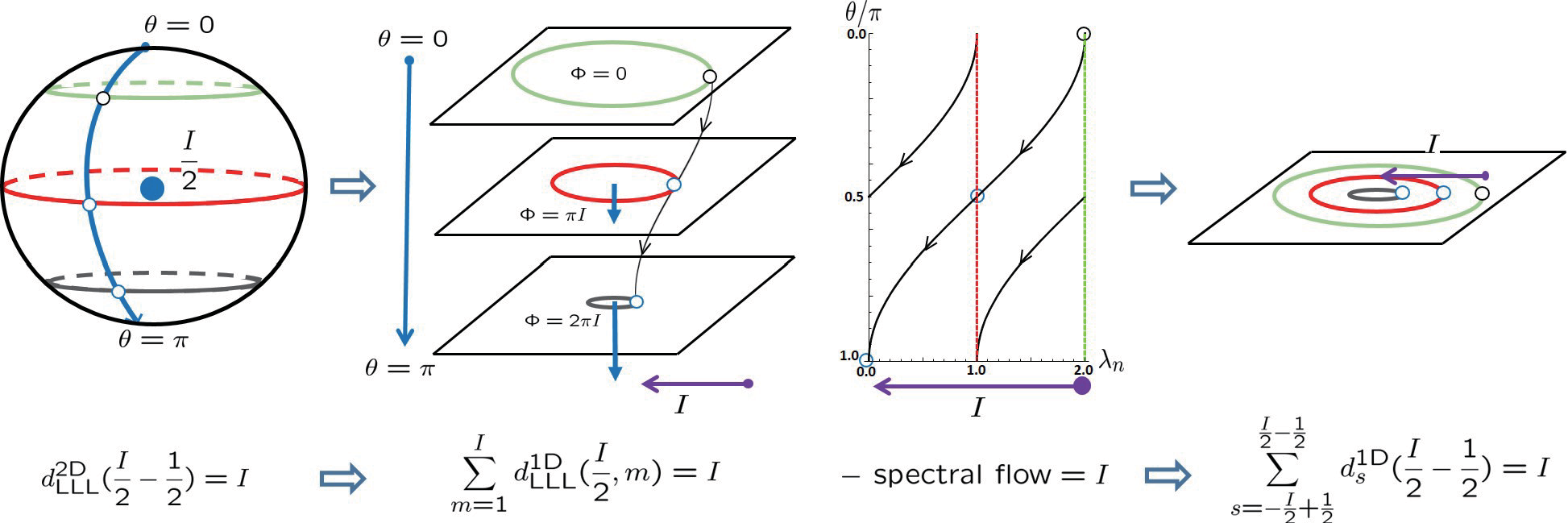}
\caption{An intuitive picture of the dimensional ladder (for $I=2$). The third figure signifies  90$^\circ$ rotated figure of evolving eigenvalues. 
}
\label{2D1D.fig}
\end{figure}

\section{BF Effective Field Theory }\label{sec:KmatrixChern-SimonsTheory}

 There arise $(2k-2)$-branes as fundamental excitations in $2k$D quantum Hall effect, and   a  $(4k-1)$D tensor  Chern-Simons action  is proposed for the description of such membranes \cite{Hasebe-2014-1}:  
\be
S=-e_{2k-2}\int_{4k-1} \mathcal{C}_{2k-1}\mathcal{J}_{2k}+\frac{\kappa}{2}\int_{4k-1}\mathcal{C}_{2k-1}\mathcal{G}_{2k}, 
\label{totalcsaction}
\ee
where 
$\mathcal{G}_{2k}=d\mathcal{C}_{2k-1}$,
$\mathcal{J}_{2k}$ denotes the $(2k-2)$-brane current, 
and $\kappa$ stands for the Chern--Simons coupling constant,
\be
\kappa
=\frac{e_{2k-2}}{2\pi} \frac{1}{m^k}.
\ee
  $m$ represents an odd integer to specify the filling factor
by $\nu_{2k}=\frac{1}{m^k}$.\footnote{$\nu_{2k}$ has nothing to do
with the winding number in Sec.~\ref{sec:generalnon-rel}.} The
tensor Chern--Simons action induces the tensor flux attachment to
membrane and accounts for non-trivial linking of $(2k-2)$-brane
currents \cite{TzeNam1989}. Let us consider  similar field theory formulation for the odd D quantum Hall effect. As the $(4k-1)$D tensor
Chern--Simons action was used for the description of the $2k$D
quantum Hall effect, one may be tempted to adopt $(4k-3)$D tensor Chern-Simons action for the present odd D case, 
but the $(4k-3)$D tensor Chern--Simons action identically
vanishes: In $(4k-3)$D, the tensor
Chern--Simons field has to be even rank, 
but the Chern--Simon
form of even rank tensor becomes 
\be
 \mathcal{C}_{2k-2}\wedge d\mathcal{C}_{2k-2}= -d\mathcal{C}_{2k-2} \wedge  \mathcal{C}_{2k-2}
=-\mathcal{C}_{2k-2} \wedge  d\mathcal{C}_{2k-2}
=0
\ee
on  closed manifolds.   The use of $(4k-3)$D  tensor Chern-Simons action  is thus in vain.   

Considering that the odd D quantum Hall
effect is embedded in one-dimension higher quantum Hall effect, a natural idea to construct effective field theory for the  
odd D quantum Hall effect will be to use one-dimension reduction of the  tensor Chern--Simons theory for the even D. 
Such theory is a tensor type BF theory in $(4k-2)$D that we will
pursue in this section. $(4k-2)$D space--time is the emergent space--time for $(2k-1)$D quantum Hall effect as $(4k-1)$D for $2k$D quantum Hall effect 
({Table~\ref{table:BFtheo}}). BF theory is a topological
field theory defined in arbitrary dimensions
\cite{Horowitz1989,Birminghametal1991,BlauThompson1991} and
describes fractional statistics of extended objects. For instance
in $(3+1)$D, BF theory is used to describe the linking and
fractional statistics
\cite{HorowitzSrednicki1990,Bergeronetal1994,Szabo1998,ChoMoore2011}
between  particle and vortex-string coupled to 2-rank 
antisymmetric tensor field \cite{Zee1994}. We will make use of such properties of BF theory 
in $(4k-2)$D space--time.

\begin{table}
\begin{center}
   \begin{tabular}{|c|c|c|c|c|c|c|c|c|c|c|c|c|}\hline
 The original space-time D  & ~2~ & ~4~ & ~6~ & ~8~  & 10 & 12 & 14 & 16  & 18 &  $\cdots$ \\ \hline
 The emergent space-time D      & ~2~ & ~6~  & 10   & 14 & 18 & 22 & 26 & 30 & 34 &  $\cdots$ \\ \hline
    \end{tabular}       
\end{center}
\caption{ 
The emergent $(4k-2)$D  space-time for the  $(2k-1)$D quantum Hall effect. } 
\label{table:BFtheo}
\end{table}

\subsection{BF theory as a reduced Chern-Simons theory}
 
 Accompanied with  one dimension reduction of the $(4k-1)$D Chern-Simons action,   
  the $(2k-1)$ tensor Chern-Simons field and $(2k-2)$-brane current $\mathcal{J}_{2k}$ in  Eq.(\ref{totalcsaction}) are decomposed as   
\begin{align}
&\mathcal{C}_{2k-1}=C_{2k-1}+C_{2k-2} dx^{4k-2}, 
\nn\\ 
&\mathcal{J}_{2k}=J_{2k-1}\wedge dx^{4k-2} +J_{2k} \label{decompcurrent}. 
\end{align}\label{decompall} 
In component representation, Eq.(\ref{decompall}) is expressed as 
\begin{align}
&C_{a_1  a_2  \cdots  a_{2k-1}} ~\rightarrow ~C_{\mu_1 \mu_2 \cdots \mu_{2k-1}}, ~C_{\mu_1 \mu_2 \cdots \mu_{2k-2}, {4k-2}}, \nn\\
&J_{a_1  a_2  \cdots  a_{2k-1}} ~\rightarrow ~J_{\mu_1 \mu_2 \cdots \mu_{2k-1}}, ~J_{\mu_1 \mu_2 \cdots \mu_{2k-2}, {4k-2}}, 
\end{align}
with $a_1, a_2, \cdots, a_{2k-1}=0, 1, 2, \cdots, 4k-2$ and $\mu_1, \mu_2, \cdots, \mu_{2k-1}=0, 1, 2, \cdots, 4k-3$. Here, $C_{2k-1}$ and $C_{2k-2}$ are taken to be independent on the compactified space coordinate $x^{4k-2}$. With $S^1$-compactification with radius $R$,  
 we derive the reduced Chern-Simons action in $(4k-2)$D as\footnote{ 
In component representation, the decomposition (\ref{decompall}) is given as follows. With  
\begin{align}
&\mathcal{C}_{2k-1}=\frac{1}{(2k-1)!} C_{a_1 a_2 \cdots a_{2k-1}} dx^{a_1} dx^{a_2} \cdots  dx^{a_{2k-1}}, \nn\\
&\mathcal{J}_{2k}=\frac{1}{(2k-1)! (2k)!} \epsilon^{a_1 a_2 \cdots a_{4k-1}} {J}_{a_1 a_2 \cdots a_{2k-1}} dx_{a_{2k}} dx_{a_{2k+1}} \cdots dx_{a_{4k-1}},  
\end{align}
we have 
\begin{align}
&C_{2k-1}=\frac{1}{(2k-1)!}C_{\mu_1\mu_2\cdots\mu_{2k-1}} dx^{\mu_1} dx^{\mu_2}\cdots dx^{\mu_{2k-1}}, \nn\\
&C_{2k-2}=\frac{1}{(2k-2)!} C_{\mu_1\mu_2\cdots\mu_{2k-2}, 4k-2 } dx^{\mu_1} dx^{\mu_2}\cdots dx^{\mu_{2k-2}}, 
\end{align}
and 
\begin{align}
&J_{2k-1}=\frac{1}{((2k-1)!)^2}\epsilon^{\mu_1\mu_2\cdots\mu_{4k-2}}J_{\mu_1\mu_2\cdots\mu_{2k-1}}dx_{\mu_{2k}}\cdots dx_{\mu_{4k-2}}, \nn\\
&J_{2k}=\frac{1}{(2k)!(2k-2)!} \epsilon^{\mu_1\mu_2\cdots\mu_{4k-2}}J_{\mu_1\mu_2\cdots\mu_{2k-2},4k-2}dx_{\mu_{2k-1}}\cdots dx_{\mu_{4k-2}}. 
\label{membranecurrentstens} 
\end{align}
The $(4k-2)$D reduced Chern-Simons action (\ref{diff4k-2redubfaction}) is expressed as 
\begin{align}
S_{(4k-2)\text{D}}
=\frac{2\pi R}{(2k-1)!}\int d^{4k-2} x &\biggl(-e_{2k-2}C_{\mu_1\mu_2\cdots\mu_{2k-1}}J^{\mu_1\mu_2\cdots\mu_{2k-1}}-e_{2k-2}(2k-1)C_{\mu_1\mu_2\cdots \mu_{2k-2}}J^{\mu_1\mu_2 \cdots \mu_{2k-2}}\nn\\
&+\frac{\kappa}{(2k-2)!}\epsilon^{\mu_1\mu_2\cdots \mu_{4k-2}}C_{\mu_1\mu_2\cdots\mu_{2k-1}}\partial_{\mu_{2k}}C_{\mu_{2k+1}\cdots \mu_{4k-2}}  \biggr)+(\text{surface-term}).
\label{chernsimons6dreduction}
\end{align}
}: 
\begin{align}
S_{(4k-2)\text{D}}&=  -2\pi R \cdot\biggl(e_{2k-2}  \int_{4k-2}(C_{2k-1}J_{2k-1}+C_{2k-2}J_{2k})-\frac{\kappa}{2}\int_{4k-2}(C_{2k-1}G_{2k-1}+C_{2k-2}G_{2k})\biggr) \nn\\
&=-2\pi R\cdot\biggl(e_{2k-2}\int_{4k-2}(C_{2k-1}J_{2k-1}+C_{2k-2}J_{2k})-{\kappa}\int_{4k-2}C_{2k-1}dC_{2k-2}+(\text{surface~term})\biggr),  
\label{diff4k-2redubfaction}
\end{align}
where  
\be
G_{2k-1}=dC_{2k-2},~~~~G_{2k}=dC_{2k-1}. 
\ee
The two tensor gauge fields, 
$C_{2k-1}$ and $C_{2k-2}$, are  coupled to different membranes with dimensions $(2k-2)$ and $(2k-3)$,  and the membrane currents are given by $J_{2k-1}$ and $J_{2k}$.   Thus there exist two kinds of membranes  in odd dimensional quantum Hall effect. However, since both $(2k-2)$ and $(2k-3)$-branes originate from  $(2k-2)$-brane in one-dimension higher space, their coupling constants to the Chern-Simons fields are equally given by $(2\pi R)e_{2k-2}$.  From Eq.(\ref{diff4k-2redubfaction}), the
equations of motion are derived as
\be
J_{2k-1}=\frac{\kappa}{e_{2k-2}}G_{2k-1}, ~~~~~J_{2k}=-\frac{\kappa}{e_{2k-2}}G_{2k}. 
\label{eqofmotresjandj}
\ee
As component equations,  Eq.(\ref{eqofmotresjandj}) becomes   
\be
J^{i_1\cdots i_{2k-2} 0}=\frac{\kappa}{e_{2k-2}}B^{i_1\cdots i_{2k-2}}, ~~~~J^{i_1\cdots i_{2k-3} 0}=\frac{\kappa}{e_{2k-2}} B^{i_1\cdots i_{2k-3}} ,  
\label{generalflulxattach}
\ee
and 
\be
J^{i_1\cdots i_{2k-1}}=-\frac{1}{(2k-2)!}\frac{\kappa}{e_{2k-2}}\epsilon^{i_1\cdots i_{4k-3}}E_{i_{2k}\cdots i_{4k-3}}, 
~~~~J^{i_1\cdots i_{2k-2}}=\frac{1}{(2k-1)!} \frac{\kappa}{e_{2k-2}}\epsilon^{i_1\cdots i_{4k-3}}E_{i_{2k-1}\cdots i_{4k-3}}.  \label{currentspacial}
\ee 
Here, 
$B^{i_1\cdots i_{2k-2}}\equiv -\frac{1}{(2k-1)!}\epsilon^{i_1\cdots i_{4k-3}}G_{i_{2k-1}\cdots i_{4k-3}}$, $B^{i_1\cdots i_{2k-3}}\equiv -\frac{1}{(2k)!}\epsilon^{i_1\cdots i_{4k-3}}G_{i_{2k-2}\cdots i_{4k-3}}$, 
$E_{i_1\cdots i_{2k-2}}\equiv G_{i_1\cdots i_{2k-2} 0}$  and $E_{i_1\cdots i_{2k-1}}\equiv G_{i_1\cdots i_{2k-1} 0}$.  
Eq.{(\ref{generalflulxattach})} signifies a generalized flux
attachment to membranes, and similarly Eq.{(\ref{currentspacial})} a
generalized Hall effect. Either $(2k-2)$-brane or $(2k-3)$-brane
current does not solely satisfy the Hall effect,  $E_{i_1\cdots i_{2k-1}}J^{i_1\cdots i_{2k-1}}\neq 0$, $E_{i_1\cdots i_{2k-2}}J^{i_1\cdots i_{2k-2}}\neq 0$, but their sum does:  
\be
E_{i_1\cdots i_{2k-1}}J^{i_1\cdots i_{2k-1}}+(2k-1)E_{i_1\cdots i_{2k-2}}J^{i_1\cdots i_{2k-2}}=0. 
\label{indehalleffecta3}
\ee
These properties reflect the fact that the present BF action originates  from a single Chern-Simons action in one dimension higher space.

\subsection{Effective action for membranes and linking number}

The present BF theory in $(4k-2)$D space-time accommodates the $(2k-2)$ and $(2k-3)$-branes with     unoccupied two-dimensional  space (Table.\ref{table:membrane}). 
\begin{table}
\begin{center}
   \begin{tabular}{|c|c|c|c|c|c|c|c|c|c|c|c|c|c|}\hline
    Dim.     &  ~~~0~~~ &  ~~~1~~~ & ~~~2~~~ & $\cdots$ & $2k-2$ & $2k-1$ &$2k-2$  & $\cdots$ & $4k-5$ & $4k-4$ & $4k-3$ \\ \hline
 $M_{2k-2}$  &  $\circ$   & $\circ$   & $\circ$  & $\cdots$  & $\circ$  & $-$ & $-$ & $\cdots$ & $-$ & $-$ &      $-$                    \\ \hline 
 $M_{2k-3}$ &  $\circ$    &  $-$ & $-$ & $\cdots$ & $-$ & $\circ$  & $\circ$  & $\cdots$    & $\circ$ & $-$ &   $-$          \\ \hline  
    \end{tabular}       
\end{center}
\caption{Two non-overlapping $(2k-2)$ and $(2k-3)$-branes in $(4k-2)$D space-time. 
From the co-dimension 2, the $(2k-2)$ and $(2k-3)$-branes are regarded as two point particles.} 
\label{table:membrane}
\end{table}
In such  two dimensions,  $(2k-2)$ and $(2k-3)$-branes are regarded as  point-like objects,  and the braiding of these objects is well-defined. In this sense, the membranes obey a generalized fractional statistics. 
 With the use of Eq.(\ref{eqofmotresjandj}), we can
express the tensor fields by the associated membrane currents, and an
effective membrane--membrane interaction can be derived from
 Eq.(\ref{diff4k-2redubfaction}) as 
\be
S_{\text{eff}}=2\pi R\cdot\frac{{e_{2k-2}}^2}{\kappa}\int d^{4k-2}x \int d^{4k-2} y~\epsilon^{\mu_1 \mu_2 \cdots \mu_{4k-2}} J_{\mu_1\cdots \mu_{2k-1}}(x) \partial_{\mu_{2k}} \biggl(\frac{1}{\partial^2}\biggr)_{x, y} J_{\mu_{2k+1}\cdots \mu_{4k-2}}(y). 
\label{effeactionmembranes}
\ee
In the thin membrane limit (see Appendix \ref{appensec:bflinking} for details), the integral part becomes the linking number between the membrane trajectories,  
\begin{align}
L(M_{2k-2}\times S^1, M_{2k-3}\times S^1)&=\frac{1}{2(2k-1)! \pi^{2k-1}} ~ \oint_{M_{2k-2}\times S^1} dx^{\mu_1 \cdots \mu_{2k-1}} \nn\\
&\times \oint_{M_{2k-3}\times S^1} dy^{\mu_{2k}\cdots \mu_{4k-3}}\cdot\frac{1}{|x-y|^{4k-2}} ~ \epsilon_{\mu_1\mu_2\cdots \mu_{4k-2}}(x_{\mu_{4k-2}}-y_{\mu_{4k-2}}),  
\label{windingsn-1lin}
\end{align}
and the effective action (\ref{effeactionmembranes}) is reduced to 
\be
S_{\text{eff}}=2\pi R\cdot\frac{{e_{2k-2}}^2}{\kappa} \cdot L(M_{2k-2}\times S^1, M_{2k-3}\times S^1 ) . 
\ee
The integral part of Eq.(\ref{windingsn-1lin}) denotes the area of $S^{4k-3}$ with coordinates $\frac{x_{\mu}-y_{\mu}}{|x-y|}$ (Appendix \ref{appensec:bflinking}), and so the linking number (\ref{windingsn-1lin}) is equivalent to the winding number accompanied with   
\be
(M^{2k-2}\times S^1)\times (M^{2k-3}\times S^1)~~\longrightarrow~~S^{n-1}. 
\ee
When the membranes are spherical, we have 
\be
(S^{2k-2}\times S^1)\times (S^{2k-3}\times S^1)~~\longrightarrow~~S^{4k-3},  
\ee
indicating that the linking number is  integer.    

In the case that quantum Hall effect contains internal degrees of freedom \cite{Hasebe-2014-2}, the  BF theory is simply generalized to include $K$ matrix: 
\begin{align}
S_{{(4k-2)}\text{D}}&=-2\pi R\cdot \biggl(e_{2k-2}\int_{4k-2}(C_{2k-1}^{\alpha}J_{2k-1}^{\alpha}+C_{2k-2}^{\alpha}J_{2k}^{\alpha})+\frac{1}{2}K_{\alpha\beta}\int_{4k-2}(C_{2k-1}^{\alpha}G_{2k-1}^{\beta}+C_{2k-2}^{\alpha}G_{2k}^{\beta})\biggr), 
\label{diff4k-2redubfactionKmat}
\end{align}
and the effective action of the membrane-membrane interaction is derived as 
\be
S_{\text{eff}}=2\pi R\cdot {e_{k-2}}^2 \biggl(\frac{1}{K}\biggr)_{\alpha\beta} L^{\alpha\beta}  , 
\ee
with 
\be
L^{\alpha\beta}\equiv L(M_{2k-2}^{\alpha}\times S^1, M_{2k-3}^{\beta}\times S^1). 
\ee




\section{Extended Dimensional Hierarchy}\label{sec:dimensionalladder}

In $2k$D quantum Hall effect, the  filling factor exhibits a hierarchical structure referred  to as the dimensional hierarchy \cite{Hasebe-2014-1,HasebeKimura2003}:   
\be
\nu_{2k}=\frac{1}{m^{\frac{1}{2}k(k+1)}}=\frac{1}{m} \frac{1}{m^2} \cdots \frac{1}{m^k} 
\label{scalingfilling2k}
\ee
or 
\be
 \nu_{2k}=\frac{1}{m^{k}}\nu_{2k-2}. 
\label{evendimqhfilling}
\ee
According to the Haldane-Halperin hierarchy
\cite{Haldane1983,Halperin1984}, in the usual 2D quantum Hall liquid,  hierarchical nature of the filling factor is a reflection of  new formations of incompressible liquids  by quasi-particle condensation.    
Therefore,  the dimensional hierarchy suggests  ``dimensional condensation'' of membrane-like excitations:   Low dimensional membranes condense to form  incompressible liquid in two-dimension higher space, and the resulting higher D  liquids also behave as membranes and condense again to form  a new  incompressible liquid in even higher dimensional space.     Such condensation process repeatedly occurs from 0 to  $2k$D,  and consequently the filling factor bears a hierarchical structure in dimensions {(\ref{scalingfilling2k})}. The dimensional hierarchy can be understood by simple dimension counting. In $2k$D quantum Hall effect there are $(2k-2)$-branes  occupying an area  ${\ell_B}^{2k}$ $(\ell_B=\sqrt{I/R})$, and the number of $(2k-2)$-branes on $2k$-sphere is evaluated as 
\be
\frac{R^{2k}}{{\ell_B}^{2k}} =I^{k}.  
\ee
Since the membrane condensation repeatedly occurs from $0$-branes to $(2k-2)$-branes,  the total number of 0-branes on $S^{2k}$ is       
\be
d_{2k}(I) \equiv I\cdot I^2 \cdots I^{k}=I^{\frac{1}{2}k(k+1)}.    
\label{d2kpoweri}
\ee
The filling factor is scaled inversely in terms of the magnetic field (or the monopole charge), and then the scaling of the filling factor becomes Eq.(\ref{scalingfilling2k}). 
One may also see that Eq.(\ref{d2kpoweri}) can be  obtained  from the lowest Landau level degeneracy of the Dirac-Landau operator, $i.e.$,  the Dirac operator index:  
\be
\text{Ind }(-i\fsl{D})_{SO(2k+1)} =d_{SO(2k+1)}[0, \frac{I}{2}-\frac{1}{2}]=\prod_{l=1}^k\prod_{i=1}^l\frac{I+l+i-2}{l+i-1} ~\sim ~ I^{\frac{1}{2}k(k+1)}. 
\ee

Let us consider whether such dimensional hierarchy will occur in the odd D quantum Hall effects. Membranes on $S^{2k-1}$ are supposed to
occupy the area ${\ell_B}^{2k-1}$, and similar dimension counting suggests 
that the number of the membranes would be  
\be
\frac{R^{2k-1}}{{\ell_B}^{2k-1}} =I^{k-\frac{1}{2}},   
\ee
and the number of 0-branes would also be 
\be
I^{\frac{1}{2}}\cdot I^{\frac{3}{2}} \cdots I^{k-\frac{1}{2}}=I^{\frac{1}{2}k^2},  
\ee
which however does not coincide with the sub-band degeneracy of the
$(2k-1)$D lowest Landau level:
\begin{align}
d_{SO(2k)}[0, \frac{I}{2}-\frac{1}{2}, s] &= \prod_{j=1}^{k-1} \frac{(I+2j-1)^2 -4s^2}{(2j)^2}\prod_{l=1}^{k-2}\prod_{j=1}^{k-l-1}\frac{I+2l+j-1}{2l+j} \nn\\& \sim~  d_{{2k-1}}(I) \equiv  I^{\frac{1}{2}(k-1)(k+2)}~~~~~~~~~~(s<<I).
\label{scaledeges2k-1}
\end{align}
The understanding of the scaling based on the simple dimensional analysis thus fails in the case of the odd D quantum Hall effects. However, Eq.(\ref{scaledeges2k-1}) suggests the alternative scaling of the  filling factor  
\be
\nu_{2k-1}=\frac{1}{m^{\frac{1}{2}(k-1)(k+2)}},  
\label{filling2k-1qhe}
\ee
 which exhibits a hierarchical structure similar to the even D case (\ref{scalingfilling2k}): 
\be
\frac{1}{m^{\frac{1}{2}(k-1)(k+2)}}= \frac{1}{m^2} \frac{1}{m^3}\cdots \frac{1}{m^k}, 
\label{filling2k-1qhedetail}
\ee
or 
\be
\nu_{2k-1}=\frac{1}{m^k}\nu_{2k-3}.  
\ee
Therefore,  the  condensation mechanism is expected to work also in the odd D quantum Hall effect.  Considering that  $(2k-1)$D quantum Hall effect is realized as a ``subspace'' of  $2k$D quantum Hall effect, it is no wonder if similar mechanism may operate in  odd D case.

From Eq.(\ref{scalingfilling2k}) and {(\ref{filling2k-1qhe})}, we find the relationship  between even D  and odd D:  
\be
\nu_{2k}= \frac{1}{m}\cdot  \nu_{2k-1}, 
\label{ladderofdimfilling}
\ee
or 
\be
I\cdot d_{{2k-1}}(I) =d_{2k}(I). 
\label{2k-12kdimlad}
\ee
Obviously, Eq.{(\ref{2k-12kdimlad})} is a manifestation of the dimensional ladder between $(2k-1)$ and $2k$D (\ref{relationroughIlarge})  discussed in Sec.\ref{sec:ladderofdim}.  
From  Eq.{(\ref{evendimqhfilling})} and {(\ref{ladderofdimfilling})}, another dimensional ladder between $(2k+1)$ and $2k$D  is inferred as 
\be
\nu_{2k+1}=\frac{1}{m^k} \nu_{2k}, 
\ee
or 
\be
I^k \cdot d_{2k}(I)=d_{2k+1}(I). 
\ee
This relation implies that $I^k$ $2k$-branes or $2k$D quantum Hall
liquids condense to form one-dimension higher $(2k+1)$D quantum Hall liquid.
{Fig.~\ref{ExtendedHierarchy.fig}} summarizes mutual relations among the
higher D quantum Hall effects that we call the extended dimensional hierarchy. 
 
\begin{figure}[tbph]\center
\includegraphics*[width=160mm]{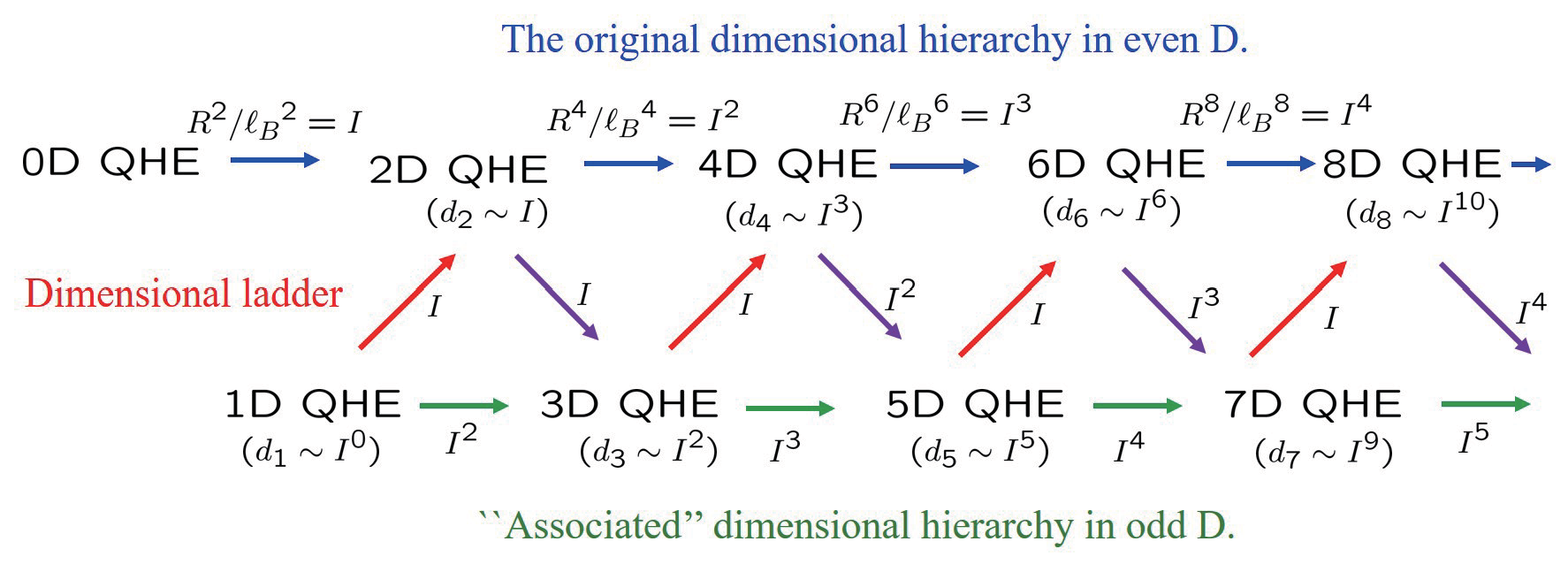}
\caption{    The extended dimensional hierarchy. The quantum Hall liquids in different dimensions are related by the web.  See also the topological periodic table (Table \ref{table:periodictable}) of A  and AIII-class.   
 }
\label{ExtendedHierarchy.fig}
\vspace{3mm}
\end{figure}

We add some comments on the relations to anomaly in relativistic
quantum field theory and D-brane physics in string theory.
The original dimensional hierarchy {(\ref{evendimqhfilling})} was only for even D quantum Hall effects, which may correspond to the 
relationship between chiral anomaly and gauge anomaly in even dimensions
\cite{Stora-1983,Zumino-1983,Baulieu-1984,GaumeGinsparg1984}. We have
pointed out a new relation in dimensions of the quantum Hall effects
{(\ref{ladderofdimfilling})} -- the dimensional ladder between even and
odd dimensions.
This ladder is associated with the relationship between even D chiral anomaly and
odd D parity anomaly \cite{AlvarezGaume-Pietra-Moore-1985}. Therefore, the 
extended dimensional hierarchy including the original hierarchy and ladder will be a counterpart of more general 
relationships among anomalies in arbitrary dimensions, such as 
Stora--Zumino chain of descent equations \cite{Stora-1983,Zumino-1983,Baulieu-1984} or the family index theorem in
differential topology \cite{Sumitani-1984,AlvarezGaume-Pietra-Moore-1985}.
On string theory side,
D-brane democracy \cite{Townsend-1995} or D-brane descent/ascent
relation \cite{AsakawaSugimotoTerashima2003} tells that any D-brane is
comprised of other D-branes in different dimensions. Obviously, such
D-brane mechanism is quite analogous to the physics of the dimensional condensation of quantum Hall liquids.

\section{Summary}\label{sec:summary}

 In this work, we developed a formulation of the odd D quantum Hall
effect in the $SO(2k-1)$ monopole background. Based on the
original Hopf maps, the chiral Hopf maps for the odd
D Landau models were introduced, and their quantization was found to give a physical realization of the 
quantum Nambu geometry. We saw that the peculiar properties of the odd D quantum Hall effect can be naturally understood, provided it is embedded in one-dimension higher even D quantum Hall
effect. We demonstrated the dimensional ladder between even and
odd D quantum Hall effects through the analysis of the
non-relativistic/relativistic Landau models by focusing on their
lowest Landau level structures. It was shown that the spectral
flow argument of the Atiyah--Patodi--Singer index theorem is the
essential mathematics behind the dimensional ladder.
From dimensional reduction of the tensor Chern--Simons theory, we
constructed BF topological field theory in emergent $(4k-2)$D
space--time for $(2k-1)$D quantum Hall effect. In the BF field
theory description, $(2k-2)$ and $(2k-3)$-branes obey a
generalized fractional statistics and bring a generalized Hall
effect and flux attachment in higher dimensions. Finally, we
proposed an extended version of the dimensional hierarchy that 
gives an overall picture about mutual relations among higher D quantum Hall effects. 

The extended dimensional hierarchy is closely related to anomaly in
relativistic quantum field theory and D-brane physics in string theory. 
We summarize the corresponding concepts in Table \ref{table:corresconcepts}. 
\begin{table}
\hspace{-0.8cm}
   \begin{tabular}{|c|c|c|c|c|c|c|c|c|}\hline
  & Fibre-bundle  & Diff. topol. & Anomaly &  NCG & QHE   \\ \hline
 $2k$D      & Chern number &  Dirac index  &  Axial anomaly    & Fuzzy even-sphere & LLL degeneracy \\ \hline
  $(2k-1)$D      & Winding number  & Spectral flow  & Parity anomaly &  Fuzzy odd-sphere & Total dim. of  LLL   \\ \hline
    \end{tabular}       
\caption{ 
The dimensional ladder appears  in apparently different theories.  } 
\label{table:corresconcepts}
\end{table}
It may be worthwhile to pursue further consequences behind such
analogies in apparently different theories.
From topological insulator point of view, the extended hierarchy is
only about the relations between A and AIII classes. The periodic
topological table suggests wider relations among topological insulators
with different symmetries. It may be interesting to see how the
extended dimensional hierarchy can further be ``extended'' to include other
topological classes. 

\section*{Acknowledgment}

 I am grateful to A. Furusaki,  J. Goryo, T. Morimoto, S. Ramgoolam, N. Sawado,   K. Shiozaki and S. Terashima for fruitful discussions and email  exchanges.   Discussions during the stay at Furusaki Riken Lab. was useful to complete the paper.   This work was supported by JSPS KAKENHI Grant No. 16K05334 and No. 16K05138. 
 
\section*{Note Added}

In completing this work, the author noticed a paper by Coskun, Kurkcuoglu and Toga that appeared on arXiv a few days ago \cite{CoskunKurkcuogluToga2016}. Their paper  has  overlaps  with the contents of Sec.\ref{sec:generalnon-rel} and Sec.\ref{landaudiracodd}.

\appendix

\section{Landau Models on $S^1$}\label{sec:reviewcircle}
 
 Here we consider Landau models on $S^1$ \cite{AsoreyEstevePacheco1983,AguadoAsoreyEsteve2001}.  Unlike other odd dimensional spheres, there does not exist the spin connection for $S^1$.  

\subsection{Non-relativistic Landau model}\label{appen:secso2landau}

We take $U(1)$ gauge field  as 
\be
A_x=-\frac{I}{2r^2}y, ~~A_y=\frac{I}{2r^2}x~~~~~~~(r^2=x^2+y^2), 
\ee
which yields the magnetic flux at the origin: 
\be
B=\partial_xA_y-\partial_y A_x= I\pi\delta(x,y). 
\ee
 Notice that a  charged particle living on $S^1$  does not classically  interact with the 
magnetic flux at the origin, but  does quantum mechanically  through the gauge potential. The Landau Hamiltonian on $S^1$ is given by  
\be
H=\frac{1}{2Mr^2}\Lambda^2,  
\label{landau1dham}
\ee
where 
\be
\Lambda=-ix(\partial_y+iA_y) + iy(\partial_x+iA_x)=-i\frac{\partial}{\partial\theta}+\frac{I}{2} . 
\label{1dcovmomentum}
\ee
We can readily solve the eigenvalue problem of the  Landau Hamiltonian. 
Due to the periodic boundary condition on $S^1$
\be
\psi(\theta=2\pi)=\psi(\theta=0),
\ee
the solutions  are classified to two cases,  even $I$ and  odd $I$.  
For even $I$, we have 
\begin{subequations}
\begin{align}
&E_n=\frac{1}{2Mr^2}n^2,\label{energyspeccir1} 
\\
&\psi_{\pm n}(\theta)=\frac{1}{\sqrt{2\pi}}e^{i\big(\pm n-\frac{I}{2}\big)\theta}, 
\label{psions1}
\end{align}
\end{subequations}
while for odd $I$, 
\begin{subequations}
\begin{align}
&E'_n=\frac{1}{2Mr^2}\left(n+\frac{1}{2}\right)^2,
\label{energyspeccir2} 
\\
&\psi'_{\pm (n+\frac{1}{2})}(\theta)=\frac{1}{\sqrt{2\pi}}e^{i\big(\pm \big(n+\frac{1}{2}\big)-\frac{I}{2}\big)\theta}, 
\label{psidashons1}
\end{align}
\end{subequations}
where $n=0, 1, 2, \cdots.$ As mentioned in Sec.\ref{sec:generalnon-rel},  for even $I$ the lowest Landau level $(n=0)$ is non-degenerate while for odd $I$  
 doubly degenerate. The energy spectrum for even $I$ (\ref{energyspeccir1}) is identical to that of the free particle on $S^1$, but the spectrum for odd $I$ (\ref{energyspeccir2}) exists only in the presence of the magnetic flux, and is 
peculiar to  quantum mechanics. 
Also notice that by the specialty of $S^1$, the magnitude of $I$ does not affect the Landau level spacing and the Landau level degeneracy. Each of the non-zero energy levels 
are two-fold degenerate corresponding to the left and right moving modes on the circle.

\subsection{Relativistic Landau model}

On the equator of $S^2$ ($\theta={\pi}/{2}$), the 1D Dirac-Landau operator (\ref{twosum1ddirac}) is given by (we have changed the basis)
\be
-i{D}_{S^1}=-i\sigma_z (\partial_\phi+i\frac{I}{2}) = 
\begin{pmatrix}
 -i\partial_{\phi}+\frac{I}{2} & 0  \\
0 &  i\partial_{\phi}-\frac{I}{2}
\end{pmatrix} =
\begin{pmatrix}
-iD_{\text{W}} & 0 \\
0 & iD_{\text{W}}
\end{pmatrix}, \label{1Dreductionofdiracindiracformalism}
\ee
where  
\be
-iD_{\text{W}}=-i\partial_{\phi}+\frac{I}{2}. 
\label{1dweyl-landau}
\ee
The Weyl-Landau operator is the covariant angular momentum itself (\ref{1dcovmomentum}), and  the $SO(2)$ Landau Hamiltonian   (\ref{landau1dham}) can be written as    
$H=\frac{1}{2M}(-iD_{\text{W}})^2.$  
For even $I$, the eigenvalues of the Dirac operator (\ref{1Dreductionofdiracindiracformalism}) are given by 
\be
\pm \lambda_n=\pm n~~~~(n=0, 1, 2, \cdots), 
\ee
and the corresponding eigenstates  are 
\be
\begin{pmatrix}
\psi_{\pm n} \\
0
\end{pmatrix},~~~~
\begin{pmatrix}
0 \\
\psi_{\mp n} 
\end{pmatrix}, 
\ee
with $\psi_{\pm n}$ (\ref{psions1}).  
Meanwhile for odd  $I$, the eigenvalues 
are given by 
\be
\pm \lambda_n 
=\pm (n+\frac{1}{2})~~~~(n=0, 1, 2, \cdots), 
\ee
and their eigenstates are 
\be
\begin{pmatrix}
\psi'_{\pm (n+\frac{1}{2})} \\
0
\end{pmatrix},~~~~
\begin{pmatrix}
0 \\
\psi'_{\mp (n+\frac{1}{2})} 
\end{pmatrix},  
\ee
with $\psi_{\pm n}$ (\ref{psidashons1}). 

The massive Dirac operator is constructed as  
\be
-iD_{S^1}+\sigma_x M = \begin{pmatrix}
-iD_{\text{W}} & M  \\
 M & iD_{\text{W}}
\end{pmatrix}, 
\label{hamiltmassive1dflux}
\ee  
which respects the chiral symmetry 
\be
\{-iD_{S^1}+\sigma_x M, \sigma_y\}=0, 
\ee
where $\sigma_y$ denotes the chiral matrix. 
The eigenvalues are readily obtained as  
\be
\pm \Lambda_{n} =\pm \sqrt{{\lambda_n}^2+M^2}~\ge M . 
\ee
For $|\Lambda_n| > M$, every Landau level is  doubly degenerate (Fig.\ref{Behaviorslambda.fig}), 
and the corresponding eigenstates are  
\begin{align}
&+\Lambda_n~(> +M):~~\frac{1}{\sqrt{2\Lambda_n(\Lambda_n+\lambda_n)}}\begin{pmatrix}
\Lambda_n+\lambda_n \\
M 
\end{pmatrix} e^{i(\lambda_n-\frac{I}{2})\phi }, ~~~~~~~~~~\frac{1}{\sqrt{2\Lambda_n(\Lambda_n+\lambda_n)}}\begin{pmatrix}
M\\
\Lambda_n+\lambda_n 
\end{pmatrix} e^{-i(\lambda_n+\frac{I}{2})\phi }, \nn\\
&-\Lambda_n~(<-M):~~\frac{1}{\sqrt{2\Lambda_n(\Lambda_n+\lambda_n)}}\begin{pmatrix}
M\\
-(\Lambda_n+\lambda_n) 
\end{pmatrix} e^{i(\lambda_n-\frac{I}{2})\phi }, ~~~~~~~\frac{1}{\sqrt{2\Lambda_n(\Lambda_n+\lambda_n)}}\begin{pmatrix}
\Lambda_n+\lambda_n \\
-M 
\end{pmatrix} e^{-i(\lambda_n+\frac{I}{2})\phi }.
\end{align}
\begin{figure}[tbph]\center
\includegraphics*[width=140mm]{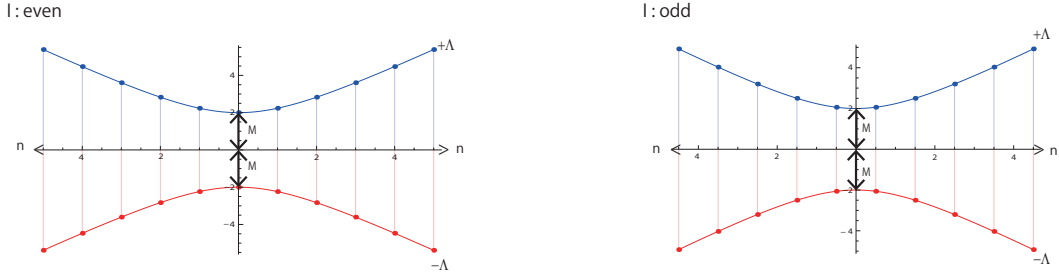}
\caption{Behaviors of  $\pm \Lambda_n$ in terms of $n$.  The left figure is for even $I$ ($I=6, M=2$), while the right figure is for odd $I$ ($I=5, M=2$). 
}
\label{Behaviorslambda.fig}
\end{figure}
 For even $I$, non-degenerate energy levels exist for $\pm \Lambda_{n=0}=\pm M$ and the eigenstates are    
\begin{align}
&+\Lambda_{n=0}(s=0)=+M~:~
\frac{1}{\sqrt{2}}\begin{pmatrix}
1 \\
1
\end{pmatrix} e^{-i\frac{I}{2}\phi}, \nn\\
&-\Lambda_{n=0}(s=0)=-M~:~ 
\frac{1}{\sqrt{2}}\begin{pmatrix}
1 \\
-1
\end{pmatrix} e^{-i\frac{I}{2}\phi}. 
\end{align}

\section{Calculation of the Winding Number }\label{append:windingso2k}

 Two patches are needed to represent  the $SO(2k)$ monopole gauge field on $S^{2k}$. The gauge field  regular except for the south/north pole is given by 
\begin{align}
&A_N=-\frac{1}{{2(1+x_{2k+1})}}\sigma_{\mu\nu}x_{\nu}dx_{\mu
}=-i\frac
{1}{2}(1-x_{2k+1})g^{\dagger} dg
, \nn\\
&A_S=-\frac{1}{{2(1-x_{2k+1})}}\bar{\sigma}_{\mu\nu}x_{\nu}dx_{\mu
}=-i\frac{1}{2}(1+x_{2k+1})gdg^{\dagger},
\end{align}
where $\sigma_{\mu\nu}$ and $\bar{\sigma}_{\mu\nu}$ are the $SO(2k)$
generators in the Weyl representation {(\ref{blockdiagso2kge})}. $g$
denotes the $SO(2k)$ gauge group element   
\be
g=\frac{1}{\sqrt{1-{x_{2k+1}}^2}} (x_{2k}-i\gamma_i x_i), 
\ee
which acts as the transition function for the $SO(2k)$ monopole gauge bundle.  
Obviously, $A_N$ and $A_S$  are related as 
\be
A_N=g^{\dagger}A_Sg-ig^{\dagger}dg.  
\ee
On the equator of $S^{2k}$, $g$ is given by 
\be
g =x_{2k} -i\gamma_ix_i ~~\in ~SO(2k), 
\ee
with 
\be
{x_i}^2+{x_{2k}}^2=1~~:~S^{2k-1}. 
\ee
 $g$ specifies a map from $S^{2k-1}$ to $SO(2k)$ and the associated winding number is  
\be
\nu_k=\frac{1}{(2k-1)! (2i)^{k-1}\mathcal{A}(S^{2k-1})}\int_{S^{2k-1}} \tr (-ig^{\dagger} dg)^{2k-1} . 
\label{windingexplicit}
\ee
We explicitly calculate Eq.(\ref{windingexplicit}) as  follows. 
Around the north-pole of $S^{2k-1}$ $(x_{2k}\simeq 1)$, $g$ behaves as 
\be
-ig^{\dagger}dg \simeq -\gamma_idx_i, 
\ee
and from the $SO(2k)$ symmetry, the integration part is represented as 
\begin{align}
\int_{S^{2k-1}} \tr (-ig^{\dagger}dg)^{2k-1} &\sim -\int_{S^{2k-1}}\tr(\gamma_i dx_i)^{2k-1} = -\int_{S^{2k-1}} dx_{1}\wedge  dx_{2} \wedge \cdots \wedge dx_{{2k-1}} \tr (\gamma_{[1} \gamma_2 \cdots \gamma_{2k-1]})\nn\\
&=\mathcal{A}(S^{2k-1})~ \tr (\gamma_{[1} \gamma_2 \cdots \gamma_{2k-1]}),   
\end{align}
where  the direction of the wedge product is taken in the inward of $S^{2k-1}$;  $\int_{S^{2k-1}} dx_{1}\wedge  dx_{2} \wedge \cdots \wedge dx_{{2k-1}} =-\mathcal{A}(S^{2k-1})$.  Eq.(\ref{windingexplicit}) now becomes 
\be
\nu_k=\frac{1}{(2i)^{k-1} (2k-1)! } \tr (\gamma_{[1} \gamma_2 \cdots \gamma_{2k-1]}), 
\label{windingexplicit2}
\ee
and then the calculation of the winding number is boiled down to taking the trace of the product of the $SO(2k-1)$ gamma matrices.  
The $SO(2k-1)$ gamma matrices in the totally symmetric representation $[0,  \frac{I}{2}]_{SO(2k-1)}$ satisfies 
\be
\gamma_{i_{2k-1}} =(-i)^{k-1} \frac{I!!}{(2k-2)!! (I+2k-4)!!}\epsilon_{i_1 i_2 \cdots i_{2k-1}} \gamma_{i_1}\gamma_{i_2}\cdots \gamma_{i_{2k-2}}~~~~(k\ge 2), 
\ee
and hence the trace is reduced to  
\begin{align}
\tr (\gamma_{[1} \gamma_2 \cdots \gamma_{2k-1]}) &=(2k-1) ~\tr (\gamma_{[1} \gamma_2 \cdots \gamma_{2k-2]}\gamma_{2k-1}) \nn\\
&= i^{k-1}\frac{(2k-1)!{(I+2k-4)}!!}{(2k-3)!!I!!}\tr({\gamma_{2k-1}}^2). 
\label{tracesum}
\end{align}
Since the $SO(2k-1)$ chiral matrix 
is given by  the following diagonal form 
\be
\gamma_{2k-1}=\text{diag}({I}_{d[0, \frac{I}{2}, \frac{I}{2}]_{SO(2k-2)}} ,  (I-2)_{d[0, \frac{I}{2}, \frac{I}{2}-\frac{1}{2}]_{SO(2k-2)}}, \cdots, (2s)_{d[0, \frac{I}{2}, s]_{SO(2k-2)}},\cdots,    
-I_{d[0, \frac{I}{2}, \frac{I}{2}]_{SO(2k-2)}}), 
\ee
with 
$I_{d[0, \frac{I}{2},  s]_{SO(2k-2)}}\equiv (\overbrace{I, I, \cdots, I}^{d[0, \frac{I}{2}, s]_{SO(2k-2)}})$,  Eq.(\ref{tracesum}) can be expressed as 
\be
\tr (\gamma_{[1} \gamma_2 \cdots \gamma_{2k-1]})=i^{k-1}\frac{(2k-1)!{(I+2k-4)}!!}{(2k-3)!!I!!}\sum_{s=-\frac{I}{2}}^{\frac{I}{2}} (2s)^2 d[0, \frac{I}{2}, s]_{SO(2k-2)},    
\ee
and the winding number (\ref{windingexplicit}) is eventually derived as 
\be
\nu_k=\frac{1}{2^{k-1}} \frac{{(I+2k-4)}!!}{(2k-3)!!I!!}\sum_{s=-\frac{I}{2}}^{\frac{I}{2}} (2s)^2 d[0, \frac{I}{2}, s]_{SO(2k-2)}    ~~~~(k\ge 2). 
\ee

\section{Weyl-Landau Problem on $S^{2k-1}$}\label{sec:weyllandaudet}

 Several properties  of the eigenstates of the Weyl-Landau operator (\ref{WeylLandauopedef}) are summarized here.  
 
\subsection{Odd $I$}

The minimum eigenvalue of the square of the Weyl-Landau operator is zero 
\be
\lambda^-_{n=0}(s=0)=0, 
\ee
for the representation  
\be
[n=0, \frac{I}{2}-\frac{1}{2}, s=0]_{SO(2k)} . 
\ee
The number of the zero-modes  is 
\be
d[0, \frac{I}{2}-\frac{1}{2}, 0]_{SO(2k)}
=\biggl(\frac{(I+2k-3)!!}{(I-1)!!(2k-2)!!} \biggr)^2 \times  \prod_{l=1}^{k-2}
\frac{(2l)!(I+l+k-2)!}{ (l+k-1)! {(I+2l-1)}! }~\sim ~I^{\frac{1}{2}(k+2)(k-1)}. 
\label{relationindextheos}
\ee
For instance, 
\begin{align}
&d[0, \frac{I}{2}-\frac{1}{2}, 0]_{SO(2)}=1=I^0, \nn\\
&d[0, \frac{I}{2}-\frac{1}{2}, 0]_{SO(4)}= \frac{1}{4}(I+1)^2~\sim ~I^2, \nn\\
&d[0, \frac{I}{2}-\frac{1}{2}, 0]_{SO(6)}=\frac{1}{192}(I+1)^2(I+2)(I+3)^2~\sim ~I^5, \nn\\
&d[0, \frac{I}{2}-\frac{1}{2}, 0]_{SO(8)}=\frac{1}{138240}(I+1)^2(I+2)(I+3)^3(I+4)(I+5)^2~\sim ~I^9. 
\end{align}

\subsection{Even $I$ ($I\neq 0$)}

The minimum eigenvalue of the square of the  Weyl-Landau operator is given by 
\be
{\lambda^-_{n=0}(s=\frac{1}{2})}^2={\lambda^-_{n=0}(s=-\frac{1}{2})}^2=(\frac{1}{2})^2,  
\ee
for the  representations, 
\be
[n=0, \frac{I}{2}-\frac{1}{2}, s=\frac{1}{2}]_{SO(2k)} 
,~~~~~[n=0, \frac{I}{2}-\frac{1}{2}, s=-\frac{1}{2}]_{SO(2k)}, 
\ee
The corresponding degeneracy is  
\begin{align}
2 ~d[0, \frac{I}{2}-\frac{1}{2}, \frac{1}{2}]_{SO(2k)}&=2~ d[0, \frac{I}{2}-\frac{1}{2}, -\frac{1}{2}]_{SO2(k)}\nn\\
&= 2 \prod_{j=0}^{k-2} \frac{(I+2j)(I+2j+2)}{(2j+2)^2}\times \prod_{l=1}^{k-2}\prod_{j=1}^{k-l-1}\frac{I+2l+j-1}{2l+j}     ~\sim~
I^{\frac{1}{2}(k+2)(k-1)}. 
\end{align}
For instance, 
\begin{align}
&d[0, \frac{I}{2}-\frac{1}{2}, \frac{1}{2}]_{SO(2)}=1-\delta_{I, 0}~\sim~I^0, \nn\\
&d[0, \frac{I}{2}-\frac{1}{2}, \frac{1}{2}]_{SO(4)}= \frac{1}{4}I(I+2)~\sim ~I^2, \nn\\
&d[0, \frac{I}{2}-\frac{1}{2}, \frac{1}{2}]_{SO(6)}=\frac{1}{192}I(I+2)^3(I+4)~\sim ~I^5, \nn\\
&d[0, \frac{I}{2}-\frac{1}{2}, \frac{1}{2}]_{SO(8)}=\frac{1}{138240}I(I+2)^3(I+3)(I+4)^3(I+6)~\sim ~I^9. 
\end{align}
Though these states are non-zero modes, in the absence of the external gauge field, these states vanish just like the zero-modes of even-sphere case.\footnote{In low dimensions, the numbers of zero-modes of the $2k$D relativistic Landau models are given by   
\begin{align}
&\text{Ind}(-i\fsl{D})_{SO(3)}=I,  \nn\\
&\text{Ind}(-i\fsl{D})_{SO(5)}=\frac{1}{6}I(I+1)(I+2), \nn\\
&\text{Ind}(-i\fsl{D})_{SO(7)}=\frac{1}{360}I(I+1)(I+2)^2(I+3)(I+4). 
\end{align}
}  (See also Appendix \ref{sec:reviewcircle} for $S^1$.)

\section{Chern Number and Winding Number}\label{append:wigingchern}

We demonstrate equality between the Chern number and the winding number associated with  monopole bundle on $2k$-sphere.  
Two patches that cover the north and south hemispheres are introduced to represent non-singular gauge field  on the base-manifold $S^{2k}$, and on their overlapping region, say the equator $S^{2k-1}$, the transition function $g$ is defined so as to satisfy 
\be
A_{N}=g^{\dagger}A_S g-ig^{\dagger}dg. 
\ee
Using $g$,  we will express the $k$th Chern number
\be
c_k=\frac{1}{(2k)!!\pi^k}\int_{S^{2k}}\tr F^k. 
\ee
Since $\tr F^k$ is given by total derivative of the Chern--Simons
form: 
\be
\tr F^k =dL_{\text{CS}}^{2k-1}, 
\ee
with   
\be
L_{\text{CS}}^{2k-1}[A]= k\int_0^1 dt~ \text{tr}(A(tdA+it^2A^2)^{k-1}), 
\ee
the Chern number can be rewritten as 
\begin{align}
c_k&=\frac{1}{(2k)!!\pi^k}(\int_{D^{2k}_N} \tr F^k +\int_{D^{2k}_S} \tr F^k)=\frac{1}{(2k)!!\pi^k}\int_{S^{2k-1}} (L_{\text{CS}}^{2k-1}[A_N] -L_{\text{CS}}^{2k-1}[A_S]) \nn\\
&=\frac{1}{(2k)!!\pi^k}\int_{S^{2k-1}} L_{\text{CS}}^{2k-1}[-ig^{\dagger}dg].  
\label{ckandlcs}
\end{align}
In the last equation, we used  $L_{\text{CS}}^{2k-1}[A_N] -L_{\text{CS}}^{2k-1}[A_S]= L_{\text{CS}}^{2k-1}[-ig^{\dagger}dg] + (\text{total~derivative~term})$,\footnote{
For  $k=2$, the 3D Chern-Simons form is given by 
\be
L_{\text{CS}}^3=\tr(AdA+\frac{2}{3}iA^3), 
\ee
and we have 
\be
L_{\text{CS}}^{3}[A_N] -L_{\text{CS}}^{3}[A_S]=\frac{1}{3}\tr(g^{\dagger}dg)^3+id(\tr A_s gdg^{\dagger})= L_{\text{CS}}^{3}[-ig^{\dagger}dg]+id(\tr A_s gdg^{\dagger}).  
\ee
} and  $L_{\text{CS}}[-ig^{\dagger}dg]$ is  the pure gauge Chern-Simons action:   
\begin{equation}
L_{CS}[A=-ig^{\dagger}dg]=(-i)^{k-1} \frac{k!(k-1)!}{(2k-1)!}~\text{tr}(-ig^{\dagger}dg)^{2k-1}.   
\end{equation}
Consequently, Eq.{(\ref{ckandlcs})} is represented as
\be
\nu_k= (-i)^{k-1} \frac{1}{(2\pi)^k} \frac{(k-1)!}{(2k-1)!} \int_{S^{2k-1}} \tr (-ig^{\dagger}dg )^{2k-1},  
\label{windingnumberdetail}
\ee
which is the winding number  from $S^{2k-1}$ to the transition function $g$.  
Thus it was shown that the Chern number {(\ref{ckandlcs})} is  equal to the winding number {(\ref{windingnumberdetail})} in general: 
\be
c_k =
\nu_k.  
\ee

\section{BF Theory and Linking Number }\label{appensec:bflinking}

In 3D, the Gauss linking number is defined for 0D objects with world lines  $V_1$ and $U_1$: 
\begin{align}
L(V_1, U_1)&=-\frac{1}{4\pi}\oint_{V_1} dx_{\mu} \oint_{U_1} dy_{\nu} ~\epsilon_{\mu\nu\rho}\partial_{\rho}^x\frac{1}{|x-y|}\nn\\
&= \frac{1}{4\pi}\oint_{V_1} dx_{\mu} \oint_{U_1} dy_{\nu} ~\epsilon_{\mu\nu\rho}\frac{x_{\rho}-y_{\rho}}{|x-y|^3}. 
\end{align}
In $n$D, 
$(p-1)$D and $(n-p-2)$D membranes undergo non-trivial linking:\footnote{Eq.(\ref{windingsn-1lin}) is obtained from Eq.(\ref{higherDgausslinkBF}) for $n=4k-2$ and $p=2k-1$.}   
\be
L(V_p, U_{n-p-1})=\frac{1}{p!(n-p-1)!}\int_{V_p} dx_{\mu_1\mu_2\cdots \mu_p}\int_{U_{n-p-1}}dy_{\mu_{p+1}\cdots d\mu_{n-1}}\epsilon_{\mu_1\mu_2\cdots\mu_n}\partial_{\mu_n}^x\biggl(\frac{1}{\partial^2}\biggr)_{xy}, 
\label{higherDgausslinkBF}
\ee
where  $V_{p}$ and $U_{n-p-1}$ respectively represent the world volumes of the $(p-1)$ and $(n-p-2)$-membranes and 
$\frac{1}{\partial^2}$ denotes the Green function in $n$D: 
\be
\frac{1}{\partial^2} =-\frac{1}{(n-2)\mathcal{A}(S^{n-1})}\frac{1}{r^{n-2}} ~~~~(n\ge 3), 
\label{expligreenlaplace}
\ee
which satisfies 
\be
\partial^2 \frac{1}{\partial^2}=\delta^{(n)}(r). 
\ee
Here, $\mathcal{A}(S^n)$ represents the area of $S^n$:\footnote{ 
Eq.(\ref{spheresurfacearea}) gives 
\be
\mathcal{A}(S^{2k})=\frac{2^{k+1}\pi^k}{(2k-1)!!},~~~~\mathcal{A}(S^{2k-1})=\frac{2\pi^k}{(k-1)!}. 
\ee
}
\be
\mathcal{A}(S^n)=\frac{2\pi^{\frac{n+1}{2}}}{\Gamma(\frac{n+1}{2})} . 
\label{spheresurfacearea}
\ee
Eq.(\ref{higherDgausslinkBF}) can be rewritten as 
\begin{align}
L(V_p, U_{n-p-1})&=-\frac{\Gamma(\frac{n}{2})}{2(n-2)\pi^{\frac{n}{2}}p!(n-p-1)!}\int_{V_p} dx_{\mu_1\cdots \mu_p}\int_{U_{n-p-1}} dy_{\mu_{p+1}\cdots  \mu_{n-1}}~\epsilon_{\mu_1\mu_2\cdots \mu_{n}}\partial_{\mu_n}^x\frac{1}{|x-y|^{n-2}}\nn\\
&= \frac{\Gamma(\frac{n}{2})}{2\pi^{\frac{n}{2}} p!(n-p)!}\int_{V_p} dx_{\mu_1\cdots \mu_p}\int_{U_{n-p-1}} dy_{\mu_{p+1}\cdots  \mu_{n-1}}~
\frac{1}{|x-y|^{n}}\epsilon_{\mu_1\mu_2\cdots \mu_{n}}(x_{\mu_n}-y_{\mu_n}).
\label{gausslinkingbftheoryhigherd}
\end{align}
This expression is mentioned in Ref.\cite{HorowitzSrednicki1990}.  

By integrating out the BF fields, 
we will derive the linking number (\ref{gausslinkingbftheoryhigherd}) from the BF action 
\begin{align}
S_{\tBF}&=\frac{1}{\theta}\frac{1}{p!(n-p)!}\int d^n x~\epsilon^{\mu_1\mu_2\cdots\mu_n}B_{\mu_1\cdots\mu_p}F_{\mu_{p+1}\cdots \mu_n}\nn\\
&-\frac{1}{p!}\int d^n x~J_{\mu_1 \cdots \mu_p}B^{\mu_1\cdots \mu_p}-\frac{1}{(n-p-1)!}\int d^n x~I_{\mu_1 \cdots \mu_{n-p-1}}A^{\mu_1\cdots \mu_{n-p-1}} , 
\label{BfcoupledJI}
\end{align}
where 
\be
F=dA,~~~~~H=dB. 
\ee
Equations of motion are derived as 
\begin{align}
&J_{\mu_1\mu_2\cdots \mu_p}=\frac{1}{\theta(n-p)!}\epsilon_{\mu_1\mu_2\cdots\mu_n}F^{\mu_{p+1}\cdots \mu_n}, \nn\\
&I_{\mu_1\mu_2\cdots\mu_{n-p-1}}=(-1)^{n(p+1)}\frac{1}{\theta (p+1)!}\epsilon_{\mu_1\mu_2\cdots\mu_n}H^{\mu_{n-p}\mu_{n-p+1}\cdots \mu_n}, \label{equationsofmotmemb} 
\end{align}
which suggest the generalized current conservation laws  
\begin{align}
&\partial_{\mu_i}J^{\mu_1 \cdots \mu_i \cdots \mu_p}=0 ~~~~~~~(i=1,2,\cdots,p),\nn\\
&\partial_{\mu_i}I^{\mu_1 \cdots \mu_i \cdots \mu_{n-p-1}}=0 ~~~(i=1,2\cdots, n-p-1). 
\end{align}
From Eq.(\ref{equationsofmotmemb}),   we have 
\begin{align}
&F^{\mu_1\mu_2\cdots\mu_{n-p}}=(-1)^{(n+1)p}~\theta\frac{1}{p!}\epsilon^{\mu_1\mu_2\cdots \mu_n}J_{\mu_{n-p+1}\cdots \mu_n},\nn\\
&H^{\mu_1\mu_2\cdots\mu_{p+1}}=(-1)^{p+1}~\theta\frac{1}{(n-p-1)!}\epsilon^{\mu_1 \cdots \mu_n}I_{\mu_{p+2}\cdots \mu_n}, 
\end{align}
and  in the Coulomb like gauge,  $\partial_{\mu} A^{\mu_1\cdots \mu \cdots \mu_{n-p-1}}=0$ and $\partial_{\mu} B^{\mu_1\cdots \mu \cdots \mu_p}=0$, $A$ and $B$ are given by 
\begin{align}
&A^{\mu_1\cdots \mu_{n-p-1}}=-(-1)^{n(p+1)}~\theta\frac{1}{p!}\epsilon^{\mu_1\mu_2\cdots \mu_n} \partial_{\mu_{n-p}}\frac{1}{\partial^2} J_{\mu_{n-p+1}\cdots \mu_{n}}, \nn\\
&B^{\mu_1\cdots \mu_p}=-\theta\frac{1}{(n-p-1)!}\epsilon^{\mu_1\cdots\mu_n}\partial_{\mu_{p+1}}\frac{1}{\partial^2}I_{\mu_{p+2}\cdots \mu_{n}}.
\end{align}
Substituting these expressions to the BF action (\ref{BfcoupledJI}),  we obtain an effective  membrane current-current interaction   
\be
S_{\text{eff}}=\theta\frac{1}{p!(n-p-1)!}\int d^n x \int d^n y ~
\epsilon^{\mu_1\mu_2\cdots\mu_n}J_{\mu_1\cdots \mu_p}(x) \partial_{\mu_{p+1}} \biggl(\frac{1}{\partial^2}\biggr)_{x,y} I_{\mu_{p+2}\cdots\mu_n}(y). 
\label{actionJJint}
\ee
In the thin membrane limit 
\begin{align}
&J^{\mu_1\cdots\mu_p}(x)=\int d^p\sigma~
 \frac{\partial( y^{\mu_1}, y^{\mu_2}, \cdots, y^{\mu_p})}{\partial(\sigma^0,\sigma^1,\cdots, \sigma^{p-1})} ~\delta^{(n)}(x-y(\sigma)), \nn\\
&I^{\mu_1\cdots\mu_{n-p-1}}(x)=\int d^p\sigma~
 \frac{\partial( y'^{\mu_1}, y'^{\mu_2}, \cdots, y'^{\mu_{n-p-1}})}{\partial(\sigma'^0,\sigma'^1,\cdots, \sigma'^{n-p-2})} ~\delta^{(n)}(x-y'(\sigma')), 
\end{align}
Eq.(\ref{actionJJint}) is reduced to  
\be
S_{\text{eff}}=\theta\frac{1}{p!(n-p-1)!}\int_{V_p} dx^{\mu_1\cdots\mu_p} \int_{U_{n-p-1}} dy^{\mu_{p+2} \cdots\mu_{n}} ~
\epsilon^{\mu_1\mu_2\cdots\mu_n} \partial_{\mu_{p+1}} \biggl(\frac{1}{\partial^2}\biggr)_{x,y}, 
\label{effectthinaction}
\ee
where 
\begin{align}
&dx^{\mu_1\cdots\mu_p}\equiv d\sigma^0 d\sigma^1 \cdots d\sigma^{p-1} \frac{\partial(x^{\mu_1}, x^{\mu_2},\cdots, x^{\mu_{p}})}{\partial(\sigma^0,\sigma^1,\cdots, \sigma^{p-1})}, \nn\\
&dy^{\mu_1\cdots\mu_{n-p-1}}\equiv d\sigma'^0 d\sigma'^1 \cdots d\sigma'^{n-p-2} \frac{\partial(y^{\mu_1}, y^{\mu_2},\cdots, y^{\mu_{n-p-1}})}{\partial(\sigma'^0,\sigma'^1,\cdots, \sigma'^{n-p-2})}. 
\end{align}
In  Euclidean space-time, 
we have 
\be
S_{\text{eff}}=\theta\cdot  L(V_p, U_{n-p-1}), 
\ee
where 
\be
L(V_p, U_{n-p-1})=\frac{1}{p!(n-p-1)! \mathcal{A}(S^{n-1})} \int_{V_p} dx^{\mu_1\cdots\mu_p} \int_{U_{n-p-1}} dy^{\mu_{p+1}\cdots\mu_{n-1}}
~
\epsilon_{\mu_1\mu_2\cdots \mu_n}\frac{x_{\mu_{n}}(\sigma)-y_{\mu_{n}}(\sigma')}{|x(\sigma)-y(\sigma)'|^n}, 
\label{linkingbevandu}
\ee
which is exactly equal to the linking number in higher dimensions (\ref{gausslinkingbftheoryhigherd}).  
With 
\be
z_{\mu}(\sigma,\sigma')\equiv \frac{x_{\mu}(\sigma)-y_{\mu}(\sigma')}{|x(\sigma)-y(\sigma')|}~~~~~(\sum_{\mu=1}^n z_{\mu}z_{\mu}=1~:~S^{n-1}),
\ee
the linking number (\ref{linkingbevandu}) can be expressed as 
\be
L=\frac{1}{\mathcal{A}(S^{n-1})}\cdot \frac{1}{(n-1)!}\int dz^{\mu_1}\cdots dz^{\mu_{n-1}} \epsilon_{\mu_1\cdots \mu_n} z_{\mu_n}, 
\label{windingsn-1}
\ee
where 
\be
dz^{\mu_1\cdots \mu_n}\equiv d\sigma^0 \cdots d\sigma^{p-1} d\sigma'^0\cdots d\sigma'^{n-p-1}
\frac{\partial(z^{\mu_1},z^{\mu_2}, z^{\mu_3}, \cdots, z^{\mu_{n}})}{\partial(\sigma^0,\cdots,\sigma^{p-1},\sigma'^0,\cdots,\sigma'^{n-p-1})}. 
\ee
To derive Eq.(\ref{windingsn-1}), we used the following formula about the determinant
\be
\frac{1}{p!(n-p)!}\epsilon_{\mu_1\mu_2\cdots \mu_n}\frac{\partial(z_{\mu_1},z_{\mu_2},\cdots, z_{\mu_p})}{\partial(\sigma_1,\sigma_2,\cdots,\sigma_p)}\frac{\partial(z_{\mu_{p+1}},z_{\mu_{p+2}},\cdots, z_{\mu_n})}{\partial(\sigma_{p+1},\sigma_{p+2},\cdots,\sigma_n)}=\frac{1}{n!}\epsilon_{\mu_1\mu_2\cdots\mu_n}\frac{\partial(z_{\mu_1},z_{\mu_2},\cdots, z_{\mu_n})}{\partial(\sigma_1,\sigma_2,\cdots,\sigma_n)}. 
\ee
The integral of Eq.(\ref{windingsn-1}) gives the area of $S^{n-1}$,  and then the linking number is equivalent to the winding number that takes integer values associated with  the map 
\be
V_p\times U_{n-p-1}~~\longrightarrow~~S^{n-1}. 
\label{windingfromtwotoone}
\ee



\end{document}